\documentclass[aps, twocolumn,superscriptaddress]{revtex4-1}%
\usepackage{epsfig,amssymb,amsmath,amsthm,amsfonts,amsbsy,mathrsfs}
\usepackage{graphicx}
\usepackage{color}
\usepackage{bm}
\usepackage{epstopdf}

\usepackage{tikz}
\usetikzlibrary{arrows,shapes,chains}
\newcommand*\circled[1]{\tikz[baseline=(char.base)]{
            \node[shape=circle,draw,inner sep=0.5pt] (char) {#1};}}

\definecolor{cream}{RGB}{222,217,201}

\begin{document}
\title{Dynamics in two-dimensional glassy systems of crowded Penrose kites}
\author{Yan-Wei Li}
\affiliation{State Key Laboratory of Polymer Physics and Chemistry, Changchun Institute of Applied Chemistry, Chinese Academy of Sciences, Changchun 130022, China}
\affiliation{Division of Physics and Applied Physics, School of Physical and
Mathematical Sciences, Nanyang Technological University, Singapore 637371, Singapore}

\author{Zi-Qi Li}
\affiliation{State Key Laboratory of Polymer Physics and Chemistry, Changchun Institute of Applied Chemistry, Chinese Academy of Sciences, Changchun 130022, China}

\author{Zhang-Lin Hou}
\affiliation{Key Laboratory of Systems Bioengineering (Ministry of Education),
School of Chemical Engineering and Technology, Tianjin University, Tianjin
300072, China}

\author{Thomas G. Mason}
\affiliation{Department of Chemistry and Biochemistry, University of California, Los Angeles, CA 90095 USA}
\affiliation{Department of Physics and Astronomy, University of California, Los Angeles, CA 90095 USA}

\author{Kun Zhao}
\email{kunzhao@tju.edu.cn}
\affiliation{Key Laboratory of Systems Bioengineering (Ministry of Education),
School of Chemical Engineering and Technology, Tianjin University, Tianjin
300072, China}

\author{Zhao-Yan Sun}
\email{zysun@ciac.ac.cn}
\affiliation{State Key Laboratory of Polymer Physics and Chemistry, Changchun Institute of Applied Chemistry, Chinese Academy of Sciences, Changchun 130022, China}
\affiliation{University of Science and Technology of China, Hefei, 230026, China}

\author{Massimo Pica Ciamarra}
\email{massimo@ntu.edu.sg}
\affiliation{Division of Physics and Applied Physics, School of Physical and
Mathematical Sciences, Nanyang Technological University, Singapore 637371, Singapore}
\affiliation{
CNR--SPIN, Dipartimento di Scienze Fisiche,
Universit\`a di Napoli Federico II, I-80126, Napoli, Italy
}
\date{\today}

\begin{abstract}
We investigate the translational and rotational relaxation dynamics of a crowded two-dimensional system of monodisperse Penrose kites, \textcolor{black}{in which crystallization, quasi-crystallization and nematic ordering are suppressed}, from low to high area fractions along the metastable ergodic fluid branch. 
First, we demonstrate a decoupling between both the translational and the rotational diffusion coefficients and the relaxation time: the diffusivities are not inversely proportional to the relaxation time, neither in the low-density normal liquid regime nor in the high-density supercooled regime.
Our simulations reveal that this inverse proportionality breaks in the normal liquid regime due to the Mermin-Wagner long-wavelength fluctuations and in the supercooled regime due to the dynamical heterogeneities.
We then show that dynamical heterogeneities are mainly spatial for translational degrees of freedom and temporal for rotational ones, that there is no correlation between the particles with largest translational and rotational displacements, and that different dynamical length scales characterize the translational and the rotational motion.
Hence, despite the translational and the rotational glass-transition 
densities coincide,
according to a mode-coupling fit, translations and rotations appear to decorrelate via different dynamical processes.
\end{abstract}

\maketitle

\section{Introduction}
The nature of glass transition remains an important and fundamental problem in condensed matter physics.
Frustration arising from many different sources, including polydispersity~\cite{Kawasaki_polydiperse}, dimensionality~\cite{Reichman_SE3D4D}, and shapes of particles~\cite{Han2011PRL,Kang_rod,Kob_dumbbells} to name a few, suppresses crystallization entirely or at least tremendously slows down the crystallization kinetics of many molecular and colloidal systems.
Thus, upon cooling or compression, many systems of entropically excited particles do not crystallize or otherwise show marked structural changes, but rather exhibit an impressive slow down in their dynamics, which becomes extremely sensitive to the control parameters.
This is the hallmark of the glass transition~\cite{Ediger,Berthier_review,Sastry_review,Parisi_review,AG,RFOT}.
In this context, shape frustration induced by the anisotropy of the particles is of particular interest, since it is ubiquitous in molecular systems.
Schematic lattice models for the investigation of the glass transitions have indeed exploited shape frustration to prevent crystallization~\cite{PicaCiamarra2003, Ciamarra2003PRE,Biroli2001}.
In colloidal experiments, understanding and controlling shape frustration would be highly desirable, since it could facilitate frustration without the need of mixing different particle shapes or sizes for systems in two and three spatial dimensions.
The anisotropy of the particles permits readily observable rotational dynamics, which might or might be not strongly coupled with translational dynamics \cite{OTP_92,OTP_96,Kob_dumbbells,Weeks_decoupling,Sung_Tracer}.
For instance, the investigation of the motion of a tetrahedral tracer in a dense amorphous suspension of colloidal spheres revealed a non-Gaussian translational displacement distribution, and a Gaussian rotational displacement distribution~\cite{Weeks_decoupling}.
Extensive simulations of polygonal tracers in two dimensions (2D) indicate that the congruence between the tracer shape and the structure of the ground state of the hosting medium critically affects the tracer's rotational diffusion~\cite{Sung_Tracer}.
Specifically, congruent structures promote a rotational hopping motion for the tracer which significantly suppresses rotational diffusion~\cite{Sung_Tracer}.
In suspensions of hard ellipses~\cite{Yilong_NC,Xu_ellipses} the translational and rotational glass transitions occur at the same density for small aspect ratios, at different densities for large ones~\cite{Yilong_NC}.
At small aspect ratios it is further observed a breakdown of both $D^T\propto \tau^{-1}$,
and $D^R\propto \tau^{-1}$, where $D^T$ and $D^R$
are the translational and the rotational diffusion coefficients and $\tau$ is the relaxation time.

In this work, we study the interplay between the translational and rotational dynamics through the numerical investigation of a monodisperse system of Penrose kites, \textcolor{black}{developing a model that mimics the experimental
system investigated in Ref.~\cite{Kun}. Some experimental data of Ref.~\cite{Kun} will be re-analysed and compared to our numerical findings}.
The shape frustration inherent in this system is strong enough to inhibit crystallization over experimentally accessible time scales~\cite{Kun,Zong2018}.
Different than ellipses~\cite{Han2011PRL,Yilong_NC,Xu_ellipses}, rods~\cite{Kang_rod,Wierenga_rod}, and other elongated
particles~\cite{OTP_92,OTP_96,Kob_dumbbells}, \textcolor{black}{ Penrose kites do not exhibit a nematic ordered phase, and are therefore ideal for studying translational and rotational glassy dynamics without inhibited motion in any degrees of freedom~\cite{Kun, Zong2018}.}
We first determine the phase diagram of the system, identifying the equation of state and the supercooled metastable branch.
We then investigate the dynamics in the supercooled regime via standard measures, as well as filtering out the effect of collective particle displacement using cage-relative (CR) measures~\cite{Weeks_longwave,Keim_MW,Shiba,Kawasaki_LongWL}, where the displacement of each particle is evaluated with respect to that of its close neighbors.
We find that while the translational dynamics is affected by the Mermin-Wagner long-wavelength fluctuations~\cite{Mermin}, the rotational one is not.
Specifically, when the relaxation dynamics is evaluated using the standard measures, in the simple liquid regime the diffusion coefficients do not scale as the inverse relaxation time, as we find $D^T\propto D^R \propto\tau^{-\kappa}$ with $\kappa > 1$.
Conversely $\kappa < 1$ in the supercooled regime.
When the relaxation dynamics is evaluated using CR measures, a different scenario emerges: in the normal liquid regime $\kappa_{\rm CR} = 1$, while in the supercooled regime $\kappa_{\rm CR} = 1$ in translation, and $\kappa_{\rm CR} < 1$ in rotation.
This implies that the rotational and the translational degrees of freedom relax via distinctly different physical processes, as we confirm by directly investigating the correlation map between translational and rotational displacements, as well as by showing that distinctly different dynamical correlation lengths characterize the translational and the rotational motion.

\section{Methods}\label{sec:method}
\subsection{Experiment}
\textcolor{black}{Our numerical model reproduces a previously investigated experimental one~\cite{Kun}. 
We provide here a few details about this experimental system as we will re-analyse some experimental data to validate our approach.
}
Penrose kites, which are four-sided polygons with one $144^{\circ}$ angle and three $72^{\circ}$ internal angles, are experimentally fabricated in the form of platelets using optical stepper lithography.
Each kite has two adjacent short edges of length $S = 1.8\pm0.1$ $\mu$m, and two adjacent long edges of length $L=2.9\pm 0.1$ $\mu$m (see inset in Fig.~\ref{fig:eos}(b)).
Full details on the preparation method can be found in Ref.~\citenum{Kun}.
Briefly, a dilute aqueous dispersion of kites is mixed with a dispersion of polystyrene spheres (diameter $\sim$40 nm, concentration $\sim0.9\%$ w/v, sulfate stabilized). 
This mixture is filled in a rectangular optical cuvette.
Kites sediment towards one surface of the cuvette, orient with their flat faces parallel with this surface, and are kept in plane via an anisotropic roughness-controlled depletion attraction between the faces of the kites and a proximate flat, smooth wall.
The interaction between kites in the plane is effectively hard.
Kites also experience viscous drag, where most of the dissipation occurs in the lubricating layer of aqueous solution between the faces of the kites and the cuvette's proximate surface.
Kites are concentrated slowly by slightly tilting the cuvette.
After $7$ months of waiting time for equilibrating the concentration profile of the system, high-resolution digital videos of kites along the length of the cuvette are taken using an optical brightfield microscope.
This provides measurements of the individual and collective dynamics of kites over a range of applied 2D osmotic pressures and different degrees of crowding.

\subsection{Simulation}
In the simulations each kite particle is constructed by lumping together $N_{d}$ small beads, as illustrated in Fig.~\ref{fig:eos}(b) (inset). The beads are rigidly connected and are placed so as to reproduce the experimentally measured ratio between the lengths of the short and of the long edges.
Considering that the interaction between different kites in the experiments is nearly hard, in the simulation we assume the beads of different kites to interact via a purely-repulsive potential that is very steep and has a hard cut-off in range, given by the the Weeks-Chandler-Andersen (WCA) potential~\cite{wca},
\begin{equation}
U(r)=
\begin{cases}
4\epsilon[(\sigma/r)^{12}-(\sigma/r)^{6}+C]& r<2^{1/6} \sigma\\
0& \text{otherwise}.
\end{cases}
\label{eq:potential}
\end{equation}
Here $r$ is the distance between two beads, $\epsilon$ is the interaction energy scale, $\sigma$ is the potential length scale. 
$C$ is a constant, which is chosen so that $U(2^{1/6} \sigma)=0$; so, the interaction is purely repulsive, and the force is continuous everywhere.
We remark that other forms of purely repulsive potentials would be expected to produce qualitatively similar results.
Length, time and pressure are reported in units of $\sigma$, $\sqrt{m\sigma^{2}/\epsilon}$ and $\epsilon/\sigma^{2}$.
All of the particles have the same mass $m$.
The ratio between the kite's perimeter, ${\rm Per}=2(L+S)$, and the number of particles making a kite, $N_d$, fix a typical length $l_{\rm rough}={\rm Per}/N_d$ which should be smaller than $\sigma$ for the roughness induced by the discrete representation of the kite to be negligible.
We have confirmed that the results are insensitive to this length scale as long as $l_{\rm rough}/\sigma < 0.22$, corresponding to $N_d = 24$.
In the following study, the results are presented for $N_d = 30$, for which $l_{\rm rough}/\sigma \simeq 0.18$.

We have performed molecular dynamics simulations of systems with the number of kites $N$ ranging from $841$ to $11250$ using periodic boundary conditions in the canonical ensemble. 
While the experiments are overdamped, for computational efficiency simulations are not. 
This implies that the simulations will not be able to describe the very short time dynamics of the systems.
The equations of motion are integrated via a Verlet algorithm~\cite{Allen_book}, and the temperature is fixed at $T=1.0\epsilon/k_{\rm B}$, where $k_{\rm B}$ is Boltzmann's constant, by using a Nos\'{e}-Hoover thermostat~\cite{Allen_book}. 
All simulations are performed with the GPU-accelerated GALAMOST package~\cite{Galamost}.

\textcolor{black}{The numerical control parameter is the number density $\rho$. To compare with the experiments, as well as to give a physical intuition of how crowded the system is, it is necessary to map this number density into the experimental area fraction.} 
Since particles in the coarse-grained model interact via a nearly hard potential that nevertheless retains residual softness, the size of bare particles can not be used directly to determine the numerical area fraction.
We solve this problem by resorting to the comparison of the locations $r_1$ and $r_2$ of the first two minima of the pair correlation functions $g(r)$, assuming that an experimental and a numerical system have the same area fraction when they have the same $r_2/r_1$ ratio. 
Further details are given in Figs.~\ref{fig:r1r2} and \ref{fig:size} in Appendix~\ref{sec:Mapsize}.
We validate this approach in Fig.~\ref{fig:eos}(b) comparing the numerical (full lines) and experimental (dashed lines) pair correlation functions at the same area fraction.
We observe that the heights of first four peaks of the numerical pair correlation function are slightly higher in the numerical data, which we attribute to the presence of a tiny polydispersity in the experimental system arising from the fabrication technique~\cite{Kun}, which is not taken into account in the simulations. In spite of this, the numerical and experimental pair correlation functions coincide in the shape and in peak positions, indicating that the numerical system mimics the experimental one in pair structure. 
We further support the mapping between simulations and experiments in Appendix~\ref{sec:Rangle}, Figs.~\ref{fig:angle} and \ref{fig:Snapshot}, where we show that the numerical distribution of the angle identified by the pointing direction of a kite relative to its closest neighbor mimics the experimental one well. For visualization purposes we have expanded the kite's size so that the simulated area fractions match with the experimental ones. 

\section{Results and discussion}
\subsection{Phase diagram}

\begin{figure}[h!]
 \centering
\includegraphics[angle=0,width=0.45\textwidth]{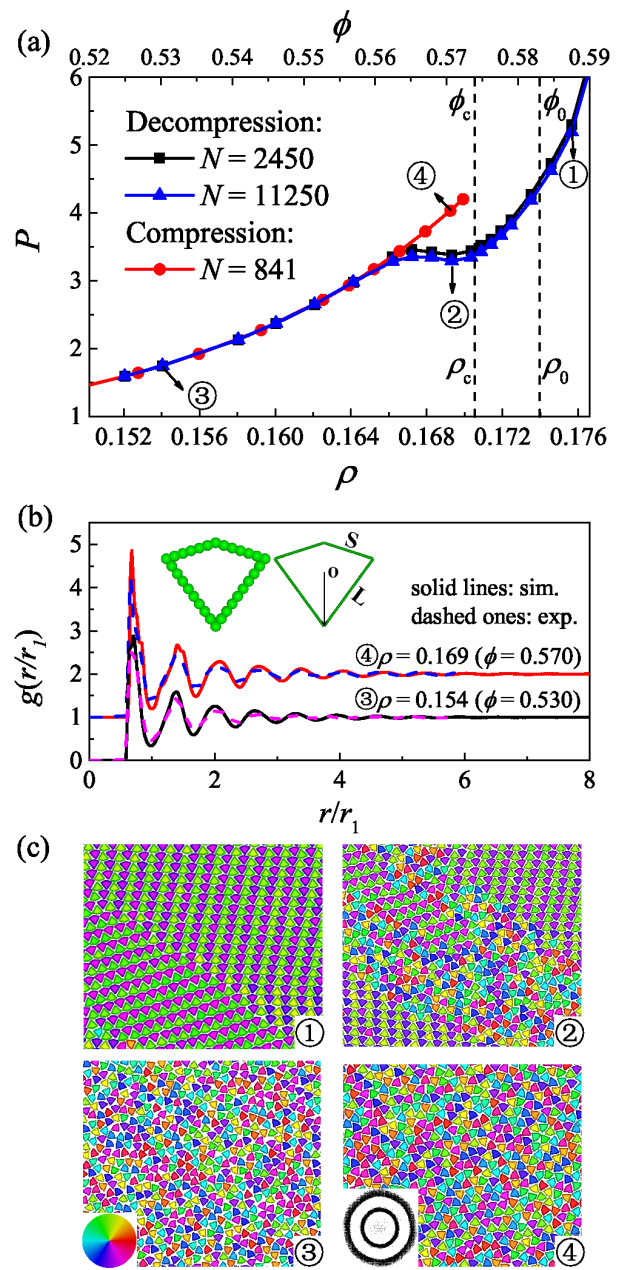}
\caption{\label{fig:eos}
(a) Equation of state of the system. The decompression (i.e. melting) curves are obtained
starting from an ASX crystalline configuration, whereas the compression curve is obtained starting from
a low-$\rho$ liquid-like configuration.
The vertical dashed lines in (a) mark the mode-coupling number density $\rho_{c}$ (area fraction $\phi_{c}$)
and the ideal glass transition number density $\rho_{0}$ (area fraction $\phi_{0}$).
(b) Scaled pair correlation functions $g(r/r_1)$ for two values of density from both simulation (full lines) and experiment (dashed lines), and illustration of the kite model.
(c) Snapshots of the system, with kites color-coded according to their pointing directions, as per the
color wheel in lower left inset, and 2D static structure factor (lower right inset). The state points for (c) are indicated in (a).
}
\end{figure}

We start by investigating the equation of state (Fig.~\ref{fig:eos}(a)) 
of the Penrose kite system, which can potentially crystallize when crowded into an alternating stripe crystal phase~\cite{Kun}, ASX, as we illustrate in Fig.~\ref{fig:eos}(c) (point \circled{1}). 
Non-local chiral symmetry, whose breaking has been found to originate from entropy in Penrose rhombs~\cite{Mason_2014SoftMatter}, is clear in the crystal phase of the kites.
This investigation of the equation of state is instrumental because it provides a value of the area fraction at which the system enters the metastable supercooled regime.
Moreover, it reveals the underlying equilibrium phases that might affect the dynamics in the supercooled regime.
In particular, since kites can tile space, one might expect their melting transition to occur via an intermediate hexatic regime~\cite{Glotzer}, as for hard hexagons~\cite{Glotzer,Kun_Hexagon}.
We numerically investigate the melting transition of the kite system by slowly decreasing the number density of a system initially in the crystalline ASX state.
Figure~\ref{fig:eos}(a) illustrates the equation of state $P(\rho)$, for two different system sizes.
Snapshots of the systems at different state points are in Fig.~\ref{fig:eos}(c), where the color code relates to the pointing direction of each kite.
The equation of state displays a Mayer-Wood loop~\cite{Mayer_wood} and a weak system size dependence, a clear indication of a first-order transition~\cite{Mayer_wood}.
Indeed, as clear from its direct visualization, the system transitions from a crystalline state at high $\rho$, Fig.~\ref{fig:eos}(c) \circled{1}, to a liquid phase with the kites pointing in random directions, \circled{3} through an intermediate crystal-liquid coexistence phase, \circled{2}.
In the crystal region, two different packing structures able to tile the space are observed, both of them with an alternating striped structure.
In one case, the pointing direction of the kites is roughly parallel to the stripe direction (Fig.~\ref{fig:eos}(c) \circled{1} upper right), while in the other case it roughly perpendicular (Fig.~\ref{fig:eos}(c) \circled{1} lower left). 
The observation of the crystal-liquid coexistence region allows us to exclude the presence of an intermediate hexatic phase~\cite{HN,KT,Y}.
We also remark that no nematic phase is observed.
Summarizing, kites melt through a first-order solid-liquid transition, as pentagons and fourfold pentilles in the family of hard polygons~\cite{Glotzer}, without an intermediate hexatic phase as in hard hexagons.
In this respect, it is interesting to notice that while kites, fourfold pentilles and hexagons can tile the plane, pentagons cannot, which indicates that tiling ability is not directly related to melting scenario.

In Fig.~\ref{fig:eos}(a) we also report the metastable equation of state of the system as obtained by slowly compressing an equilibrated very dilute configuration. 
This equation of state extends up to the maximum number density at which we are able to relax the system, just below the number density associated with the mode-coupling glass transition, $\rho_c \simeq 0.171$ ($\phi_c \simeq 0.574$).
The compression curve departs from the decompression one and extends to the supercooled regime where the equilibrium configurations are of coexistence type.
\textcolor{black}{The absence of long-range spatial order and quasi-crystalline order in the supercooled regime is apparent from the pair correlation function, illustrated in Fig.~\ref{fig:eos}(b), from the direct visualization of Fig.~\ref{fig:eos}(c) \circled{4}, and from the ring-like pattern of the static structure factor (Fig.~\ref{fig:eos}(c) \circled{4} inset).
Thus, consistent with the experimental results~\cite{Kun}, our simulations show that kites neither crystallize or quasi-crystallize nor undergo an isotropic-nematic transition under compression conditions employed by this study, as a consequence of the existence of a variety of local polymorphic configurations (LPCs), introduced in the analysis of ref.~\cite{Kun}.}
Differently~\cite{Glotzer,Mason_book,Wang_Mason,Mason_Lock_key},
hard regular polygons, such as triangles~\cite{Kun_triangles}, squares~\cite{Kun_squares},
and pentagons~\cite{Kun_Pentagons}, crystallize when compressed,
while hard ellipses, rods and dumbbells, might go through an isotropic-nematic transition,
depending on the aspect ratio~\cite{Xu_ellipses,Yilong_NC}.
This peculiarity of the Penrose kite system makes it ideal for
the investigation of the glassy dynamics.

\subsection{Supercooled dynamics}

\begin{figure}[tb]
  \centering\includegraphics[angle=0,width=0.4\textwidth]{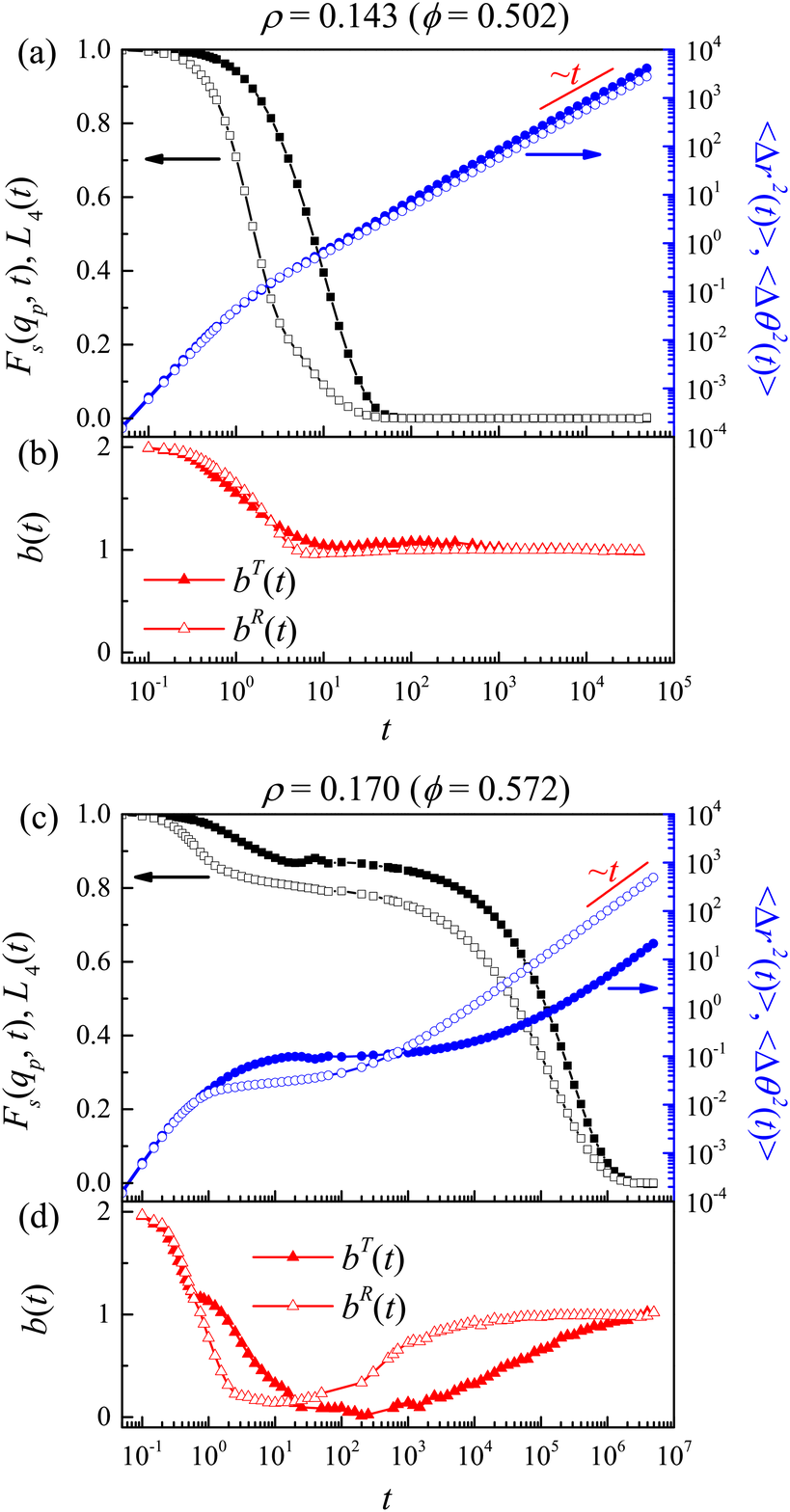}
 \caption{Characterization of the structural relaxation at (a) $\rho=0.143$ ($\phi=0.502$)
and at (b) $\rho=0.170$ ($\phi=0.572$). The figure illustrates the MSD (blue closed circles),
MSAD (blue open circles), ISF (black closed squares)and RCF (black open squares) as well as the log-slopes of the MSD $b^{T}(t)$ (red closed triangles) and of the MSAD $b^{R}(t)$ (red open triangles).
Here, the angular displacement in MSAD is in radian unit.
 \label{fig:glass}
}
\end{figure}

We now turn to the investigation of the relaxation dynamics of the kite system, which we have carried out using standard quantities:
the mean square displacement (MSD), $\langle \Delta r^{2}(t)\rangle=\langle \frac{1}{N}\sum_{j=1}^{N}[\mathbf{r}_{j}(t)-\mathbf{r}_{j}
(0)]^2\rangle$, the mean square angular displacement (MSAD), $\langle \Delta\theta^{2}(t)\rangle=\langle \frac{1}{N}\sum_{j=1}^{N}[\theta_{j}(t)-\theta_{j}(0)]^2\rangle$, the self-intermediate scattering function (ISF) $F_{s}(q_{p},t)=\langle \frac{1}{N} \sum_{j=1}^{N}e^{i\mathbf{q_{p}}\cdot(\mathbf{r}_{j}(t)-{\mathbf{r}}_{j}(0))}\rangle$, the $n$-fold rotational correlation function (RCF) $L_{n}(t)=\langle \frac{1}{N}\sum_{j=1}^{N}\cos[n (\theta_{j}(t)-\theta_{j}(0))]\rangle$, as well as the log-slopes of the MSD, 
$b^{T}(t)= \mathrm{d}\left(\mathrm{ln}\langle \Delta r ^{2}(t)\rangle\right)/\mathrm{d}(\ln(t))$
and of the MSAD, $b^{R}(t)= \mathrm{d}\left(\mathrm{ln}\langle \Delta \theta ^{2}(t)\rangle\right)/\mathrm{d}(\ln(t))$.
Here, $\mathbf{r}_{j}(t)$ and $\theta_{j}(t)$ are the position of the center of mass and the orientation of kite $j$ at time $t$, $q_{p}=|\mathbf{q_{p}}|$ is the wavenumber corresponding to the first peak of the structure factor.
\textcolor{black}{We also fix $n=4$ in our analysis~\cite{Yilong_NC}, the value at which the peak height of the susceptibility of RCF attains its maximum value, at $\rho=0.169$ ($\phi=0.570$).}
From the mean squared displacements we extract the long-time translational and rotational diffusion coefficients,
$D^T = \lim_{t \to \infty} \langle \Delta r^{2}(t)\rangle/4t$ and 
$D^R = \lim_{t \to \infty} \langle \Delta \theta^{2}(t)\rangle/2t$, respectively.
We measure the translational relaxation time $\tau^{T}$ and the rotational relaxation time $\tau^{R}$ as those corresponding to $F_{s}(q_{p},\tau^{T}) = 1/e$ and $L_{n}(\tau^{R})= 1/e$, respectively. All data presented in the following are collected after equilibrating the system for about $40\tau^{T}$.

Figure~\ref{fig:glass} illustrates the time evolution of these quantities at
low (a) or high (b) number densities. 
At high number density $F_{s}(q_{p},t)$ and $L_{n}(t)$
develop a two-step decay, and the MSD and MSAD develop a plateau, which are signatures of glassy dynamics. 
Concurrently, $b^{T}(t)$ and $b^{R}(t)$ acquire values below $1$ and even down to $\simeq 0$, highlighting the sub-diffusive relaxation dynamics. 

We evaluate the translational and rotational glass transition number densities by fitting the
$\rho$ dependence of $\tau^{T}$ and of $\tau^{R}$ with the mode-coupling power-law prediction (MCT), $\tau \sim(\rho_{c}-\rho)^{-\gamma}$~\cite{Ediger}, where
$\rho_{c}$ is the MCT glass transition point, and with the Vogel-Fulcher-Tammann
(VFT) law, $\tau \sim \exp(D_{f}\rho/(\rho_{0}-\rho))$~\cite{Ediger}, where $D_{f}$ is
the fragility index and $\rho_{0}$ is the ideal glass transition point.
We show the fitting results in Fig.~\ref{fig:fit} in Appendix~\ref{sec:dyna}.
We find the translational and the rotational relaxation time to diverge together, at
$\rho_{c}^{T}=\rho_{c}^{R}=0.171\pm0.001$ according to the MCT fit, and at
$\rho_{0}^{T}=\rho_{0}^{R}=0.174\pm0.0004$ according to the VFT fit.
These critical values are indicated with vertical dashed lines in Fig.~\ref{fig:eos}(a).
The synchronous arrest of the translational and of the rotational relaxation dynamics excludes the existence of intermediate states, i.e., rotational glass (liquid) yet translational liquid (glass), as observed in hard ellipses at certain values
of aspect ratio~\cite{Xu_ellipses,Yilong_NC}.
We also notice that the VFT fits yield $D_{f}^{T} \simeq 0.238 < D_{f}^{R} \simeq 0.361$, 
indicating that the translational dynamics is more fragile than the rotational one (see Fig.~\ref{fig:fit} in Appendix~\ref{sec:dyna}). 
This result also signals a decoupling between the translational and the rotational relaxation dynamics, which we will discuss later on.

\subsection{Cage-relative dynamics\label{sec:cr}}

\begin{figure}[h!]
 \centering
 \includegraphics[angle=0,width=0.4\textwidth]{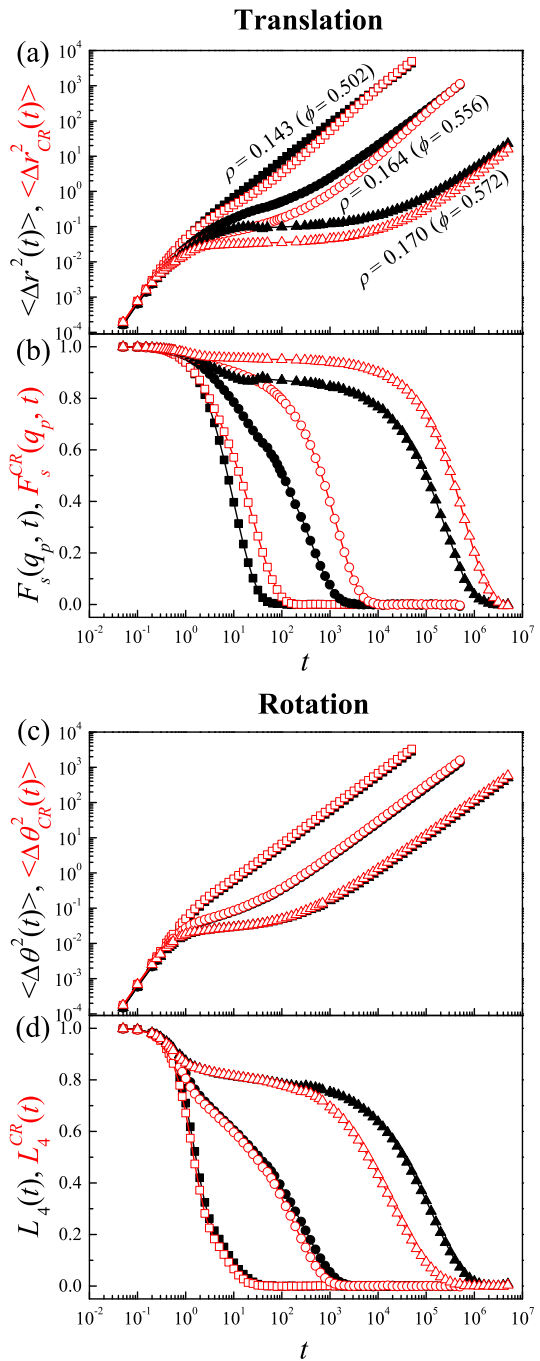}
 \caption{Translational (left column) and rotational (right column) dynamics
characterized by standard measures (black) and by CR measures (red)
for three values of the number density. (a) MSD (black) and
CR-MSD (red). (b) MSAD (black) and CR-MSAD (red).
(c) ISF (black) and CR-ISF (red).
(d) RCF (black) and CR-RCF (red).
In all panels, the number densities are $\rho$ = 0.143 (squares), 0.164 (circles), and 0.170 (triangles), and the corresponding area fractions are indicated in (a).}
  \label{fig:dyn}
\end{figure}

The translational relaxation dynamics of 2D systems has been recently shown to be strongly affected by the collective particle motion arising from the Mermin-Wagner long-wavelength fluctuations~\cite{Weeks_longwave,Keim_MW,Shiba,Kawasaki_LongWL,unpublished}.
The effect of these fluctuations can be filtered out investigating the dynamics
using CR measures, and indeed recent results have clarified~\cite{Weeks_longwave,Keim_MW,Shiba,Kawasaki_LongWL,unpublished}
that when CR quantities are considered the previously observed fundamental distinctions~\cite{Szamel_2d3d} between the glass transition in 2D and in 3D disappear.
However, it remains unclear how these fluctuations affect
the scaling relations between diffusion coefficient and relaxation time in both translation and rotation, also because all previous studies of the CR dynamics focused on systems with a radially symmetric interaction potential.
Here we investigate this issue studying the dynamics of our system
using CR quantities.
To perform the CR investigation, we determine the neighboring information via a Voronoi construction.
The CR mean square displacement (CR-MSD) is defined as
$\langle
\Delta r^{2}_{CR}(t)\rangle=\langle \frac{1}{N}\sum_{j=1}^{N}[(\mathbf{r}_{j}(t)-\mathbf{r}_{j}
(0))-
\frac{1}{N_{j}}\sum_{m=1}^{N_{j}}(\mathbf{r}_{m}(t)-\mathbf{r}_{m}(0))]^{2}\rangle$,
where the second sum runs over all $N_{j}$ Voronoi neighbors of particle $j$.
Similarly, the CR mean square angular displacement
(CR-MSAD) is $\langle
\Delta \theta^{2}_{CR}(t)\rangle=\langle \frac{1}{N}\sum_{j=1}^{N}[(\theta_{j}(t)-\theta_{j}(0))-
\frac{1}{N_{j}}\sum_{m=1}^{N_{j}}(\theta_{m}(t)-\theta_{m}(0))]^{2}\rangle$.
We also measure the CR intermediate scattering function (CR-ISF) 
and 4-fold rotational CR correlation function (CR-RCF), respectively defined as
$F_{s}(q_{p},t)=
\langle\frac{1}{N} \sum_{j=1}^{N}e^{i\mathbf{q_{p}}\cdot[(\mathbf{r}_{j}(t)-{\mathbf{r}}_{j}(0))-
\frac{1}{N_{j}}\sum_{m=1}^{N_{j}}(\mathbf{r}_{m}(t)-\mathbf{r}_{m}(0))]}\rangle$
and as
$L_{4}(t)=\langle \frac{1}{N}\sum_{j=1}^{N}\cos[4(\theta_{j}(t)-\theta_{j}(0) -
\frac{1}{N_{j}}\sum_{m=1}^{N_{j}}(\theta_{m}(t)-\theta_{m}(0)))]\rangle$.

Figure~\ref{fig:dyn} compares the standard (black full symbols) and the CR (red open symbols) dynamical quantities.
Figures~\ref{fig:dyn}(a) and \ref{fig:dyn}(b) focus on the translational dynamics, and evidence that the CR-MSD has a smaller Debye-Waller plateau value than the MSD, and that the CR-ISF has a higher plateau than the ISF. 
This is as expected, given that long-wavelength fluctuations lead to a reduction in the apparent translational confinement of the particles~\cite{Weeks_longwave,Keim_MW,Shiba,Kawasaki_LongWL}.
Figures~\ref{fig:dyn}(c) and \ref{fig:dyn}(d) focus on the rotational dynamics. In Fig.~\ref{fig:dyn}(c) we observe no
significant difference between the MSAD and CR-MSAD. Since the MSD
is dominated by the particles with the largest displacements, this result implies that
the particles with the largest rotations do not rotate collectively with their close neighbors, and are thus not affected by the CR measures.
Figure~\ref{fig:dyn}(d) reveals that there is no sensible difference between RCF and CR-RCF in the ballistic and caging stages.
Conversely, on the $\alpha$ relaxation time scale,
where vibrational modes play a minor role
as particles escape from the cages formed by their neighbors and start diffusing,
the two measures differ.
In particular, the CR rotational scattering function decays well before the standard one, which is just the opposite of what observed for the translational correlation function in Fig.~\ref{fig:dyn}(c).
We have verified that this occurs because of the existence of particles with a small rotational displacement which are close to particles with a very large one. Under this condition the CR measures consider both particles as relaxed, which leads to a faster decay of the CR correlation function with respect to the standard one. \textcolor{black}{These results indicate that long-wavelength fluctuations affect the translational dynamics, as previously observed~\cite{Weeks_longwave,Keim_MW,Shiba,Kawasaki_LongWL}, while they do not affect the rotational dynamics, as expected.}


\subsection{Decoupling between diffusion and relaxation}
\begin{figure}[tb]
 \centering
 \includegraphics[angle=0,width=0.42\textwidth]{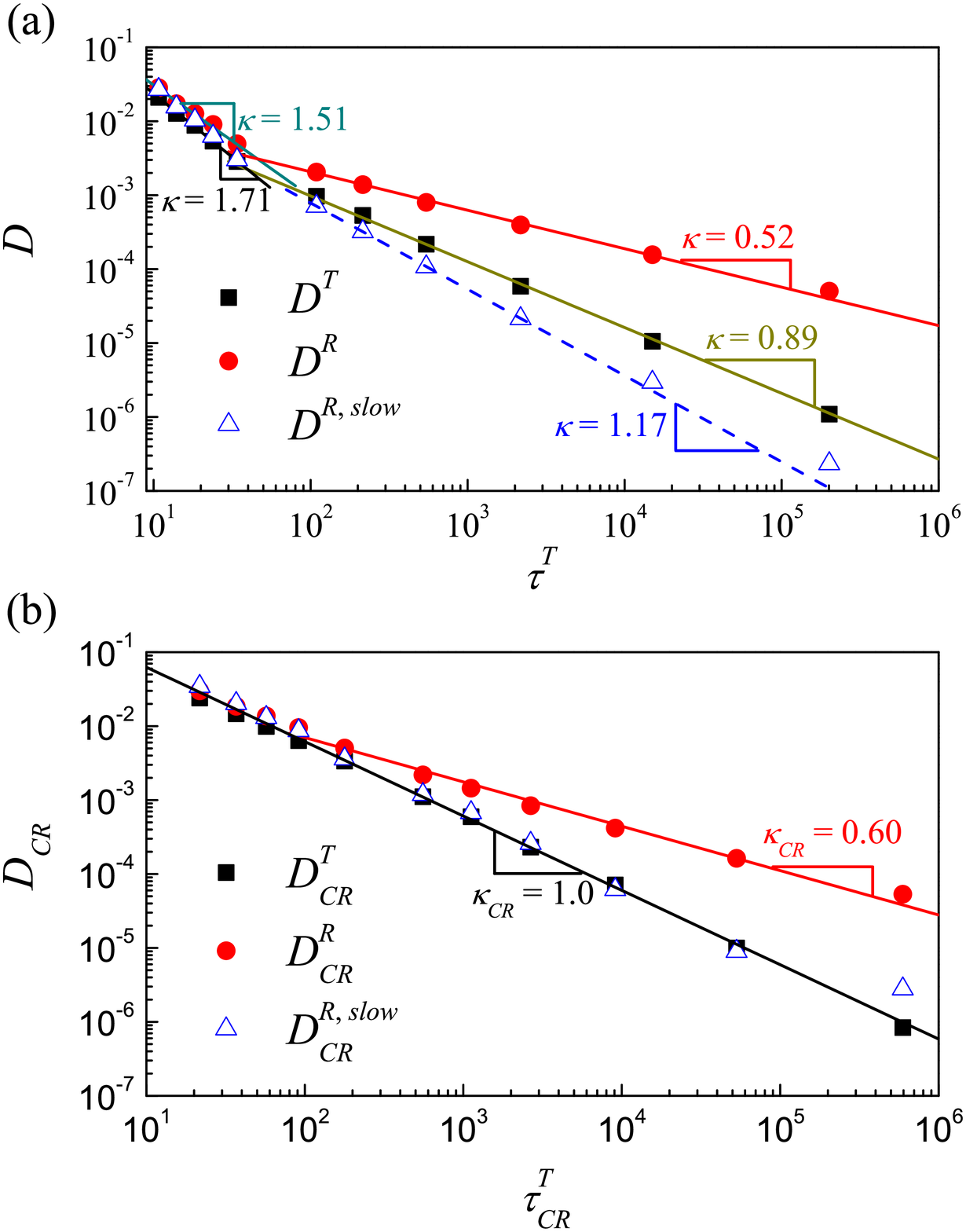}
 \caption{Dependence of the diffusion coefficient on the relaxation time $\tau^{T}$
 for both translation and rotation, evaluated using (a) standard and (b) CR measures.
 The unusual breakdown with $k > 1.0$ observed in the liquid phase using standard measures is not
 observed in the CR quantities.
  In both panels, we also plot results for the rotational diffusion coefficient of the slow  particles,
 as estimated from a Gaussian fit of their angular displacement distribution (see e.g., Fig.~\ref{fig:dtr}(b) and text).
 The solid and dashed lines in (a) and (b) mark the scaling relationship
$D\propto(\tau^{T})^{-\kappa}$.}
  \label{fig:se}
\end{figure}

The decay of the self-intermediate scattering function investigated at a wavenumber $q$
is affected by particle displacements of order $2\pi/q$.
Large amplitude vibrations, which are facilitated in the presence of
long-wavelength fluctuations, might therefore induce the decay of the translational relaxation function
without causing any structural rearrangement and hence without promoting
particle diffusion. This implies that the inverse proportionality between diffusion coefficient and relaxation time
may break down with $D\propto \tau^{-\kappa}$ with $\kappa > 1$, not with
$\kappa < 1$ as usually observed in the supercooled regime. 
Such a breakdown has been recently observed in a variety of different
2D models~\cite{SE_softdisk,Sastry_SE, SE_Ellipses,Sung_Tracer,unpublished}.
This unusual breakdown is restricted to the liquid regime because upon supercooling the vibrational amplitudes decrease, so that the breakdown of the inverse proportionality becomes dominated by the dynamical heterogeneities, which promote $\kappa < 1$.
\textcolor{black}{The crossover density at which $\kappa$ becomes $< 1$ marks the onset of the slow dynamics, for the considered system size~\cite{Sastry_SE,Sung_Tracer,Water_SE,Li_Decoupling,SE_experiment,Weeks_decoupling,Kob_dumbbells,SE_softdisk,SE_Ellipses}.}
In Fig.~\ref{fig:delta} in Appendix~\ref{sec:regime}, we also show the log-slope of the MSD evaluated at the time scale $\tau$,
$b^{T}(\tau)$, acquires its minimum value at onset density, where the system is therefore maximally sub-diffusive. 
This appears to be a general features of glassy systems~\cite{unpublished}.

For the kite system, such a crossover in the scaling parameter $\kappa$ is shown in Fig.~\ref{fig:se}(a) (black squares). 
The figure also illustrates that the rotational dynamics (red circles) undergoes an analogous crossover, but with a smaller $\kappa$ value in the supercooled regime. This suggests that the rotational dynamics is more heterogeneous than the translational one, as we will confirm with other measures later on.
One might suspect that the crossover in $\kappa$ originates from the fact that the diffusion coefficient $D$ is not evaluated at $\tau$. However, we do have checked that the same crossover occurs in the $\tau$ dependence of the effective instantaneous translational diffusion coefficient~\cite{Mason2000} evaluated at the relaxation time scale, $D(\tau^T) = \frac{1}{4}\left. \frac{d \langle \Delta r^{2}(t)\rangle}{dt}\right|_{\tau^T}$. An analogous result occurs for the rotational motion.

Figure~\ref{fig:se}(b) illustrates the dependence of the diffusion coefficient on the relaxation time, when these quantities are evaluated using the CR measures. 
In this case, $D_{CR}^{T}\propto(\tau_{CR}^{T})^{-1}$ holds both in the normal liquid and in the supercooled regime. 
Since $D_{CR} \simeq D$, the differences in $\kappa$ between the standard and the CR measures stem from the differences in the relaxation times.
Indeed, $\tau_{CR}^T > \tau^T$ as CR measures remove the effect of the long-wavelength fluctuations, which accelerate the relaxation. 
The upshot of this discussion is that the $\kappa > 1$ values 
surprisingly and ubiquitously observed~\cite{SE_softdisk,Sastry_SE, SE_Ellipses,Sung_Tracer} in the normal liquid regime
of 2D systems are due to translational long-wavelength fluctuations.

We wish to remark that, for the translational dynamics, $\kappa_{CR} =1$ in the supercooled regime, where the dynamics is heterogeneous. 
\textcolor{black}{This indicates that CR measures are less sensitive to the heterogeneous dynamics than the standard one. This is so as the heterogeneous dynamics is generally associated to the coordinated displacement of close particles, which the CR measures may filter out.}

Figure~\ref{fig:se}(b) also illustrates that there is a standard breakdown in rotation, with
$D_{CR}\propto(\tau^{T}_{CR})^{-\kappa_{CR}}$ and $\kappa_{CR} = 1$ at low area fractions,
$\kappa_{CR} < 1$ in the supercooled regime.
Hence, the breakdown of the rotational inverse proportionality between diffusion coefficient and relaxation time cannot be
attributed to the Mermin-Wagner long-wavelength fluctuations.

\begin{figure}[tb]
 \centering
 \includegraphics[angle=0,width=0.45\textwidth]{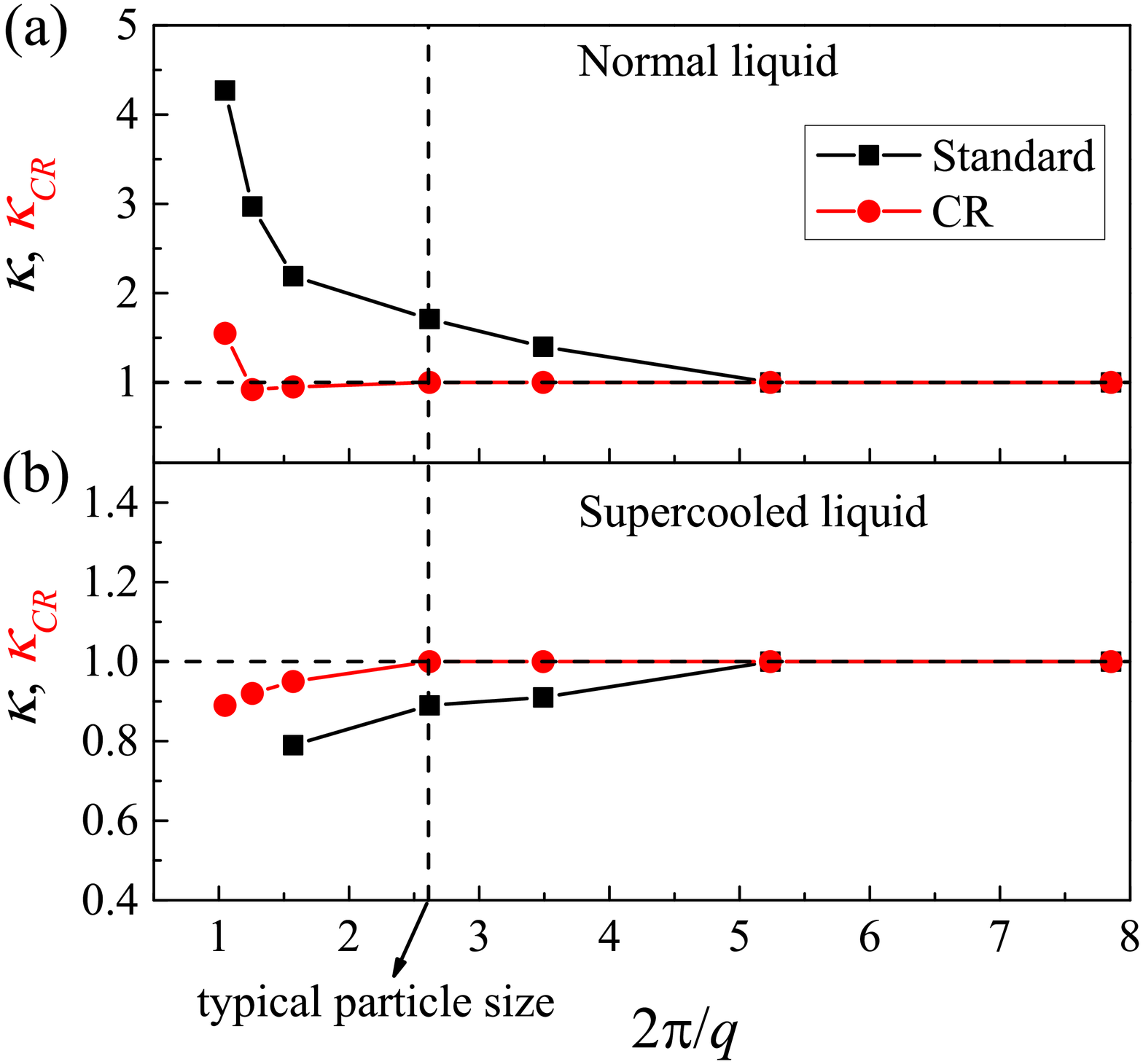}
 \caption{Length scale dependence of the exponents $\kappa$ (black squares) and $\kappa_{CR}$ (red circles)
 characterizing the breakdown of the inverse proportionality between diffusion coefficient and relaxation time for standard measures ($D \propto \tau(q)^{-\kappa}$)
 and CR measures ($D_{CR} \propto \tau_{CR}(q)^{-\kappa_{CR}}$)
 in (a) the normal liquid regime and in (b) the supercooled regime.
 The relaxation time $\tau$ ($\tau_{CR}$) is calculated from ISF (CR-ISF) with different values of $q$.}
  \label{fig:kappa}
\end{figure}

\textcolor{black}{
The ISF $F_s(q,t)$ essentially measures the fraction of particles with displacement $\Delta r(t) < 2\pi/q$ at time $t$.
The displacement of a particle can be considered as arising from two contributions, irreversible rearrangements, $\Delta r_{\rm irr}(t)$, alike particle jumps~\cite{Ciamarra2015SM}, and oscillatory motion $\Delta r_{\rm osc}(t)$. 
The role of the oscillatory motion in the decay of the ISF is generally neglected in 3d as its maximum amplitude, related the Debye-Waller factor, is commonly much smaller than the probed wavelength $2\pi/q$. 
This approximation does not hold in 2d, where long-wavelength fluctuations play a major role. 
The effect of these long-wavelength fluctuations on the decay of the ISF is however $q$ dependent. 
Indeed, at small $q$ only the longest wavelength may contribute displacements of order $2 \pi/q$ to affect the relaxation. 
However, the longest wavelength develops slowly, which implies that when the overall displacement of a particle is $2 \pi/q = \Delta r_{\rm irr}(t) + \Delta r_{\rm osc}(t)$, the overall displacement is dominated by its irreversible component, $\Delta r_{\rm irr}(t) \gg \Delta r_{\rm osc}(t)$.
This implies that the decay of the ISF at large enough $q$ is only affected by the irreversible displacements. Since these displacements are also those fixing the diffusivity of the system, at large $q$ one expects $\kappa = 1$.
}

To verify this physical picture we investigate the $q$ dependence of the scaling relation between diffusion coefficient and q-dependent relaxation time.
Figure~\ref{fig:kappa} illustrates the dependence of the scaling exponents $\kappa$, extracted from $D\propto(\tau^{T})^{-\kappa}$, and $\kappa_{CR}$ extracted from $D_{CR}\propto(\tau^{T}_{CR})^{-\kappa_{CR}}$, on the probing length scale $2\pi/q$.
The vertical dashed line at $d=2\pi/q_{p}$ corresponds to the typical particle size.

\textcolor{black}{In the normal liquid regime (see Fig.~\ref{fig:kappa}(a)),
the inverse proportionality between $D$ and $\tau^{T}$ is broken with $\kappa >1.0$ at small length scales due to the long-wavelength fluctuations,
and it is gradually recovered as the probing length scale increases,
in agreement with our expectation.
When the scaling relation between diffusivity and relaxation time is evaluated comparing the CR diffusivity, one finds $\kappa_{CR}\simeq 1.0$ in a large range of length scales.
There is a weak breakdown at very small length scales,
i.e., $\leq0.5d$, at which noise dominates the relaxation time.}

\textcolor{black}{In the supercooled liquid regime (see Fig.~\ref{fig:kappa}(b)),
$\kappa<1$ below a length scale $\xi_{\rm DH}\simeq 2d$
which measures the typical size of the dynamical heterogeneities,
and $\kappa = 1$ for larger length scales.
For the CR measures, we find $\xi_{\rm DH}^{\rm CR}\simeq d$,
further indicating that CR measures affect the estimation of the degree of correlation of the dynamics.
This is the case as the CR measures filters out the effect of
correlated particle displacements of close particles,
regardless of their physical origin.
}

\subsection{Heterogeneities and displacement distribution functions}
\begin{figure}[!tb]
 \centering\includegraphics[angle=0,width=0.47\textwidth]{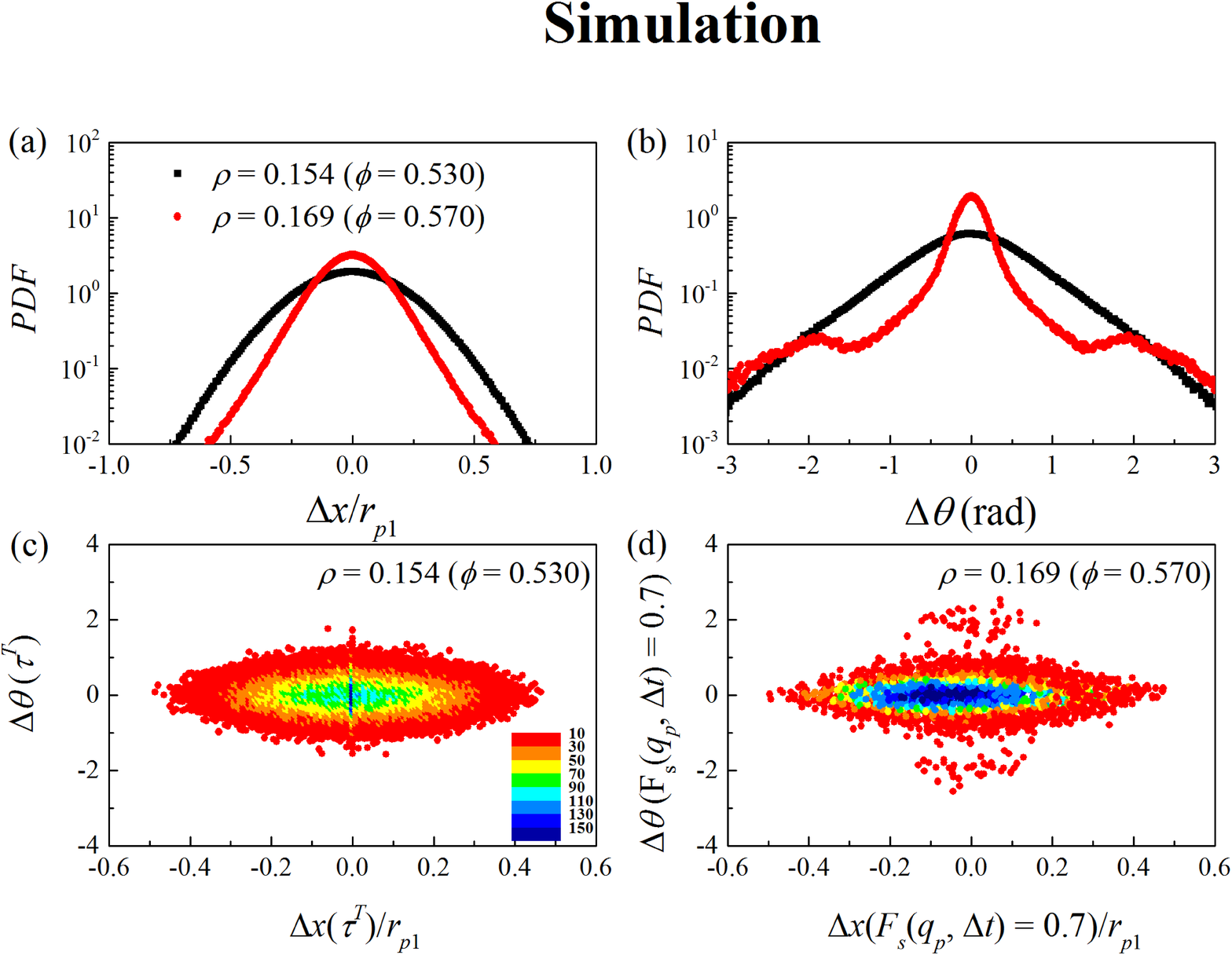}
  \centering\includegraphics[angle=0,width=0.47\textwidth]{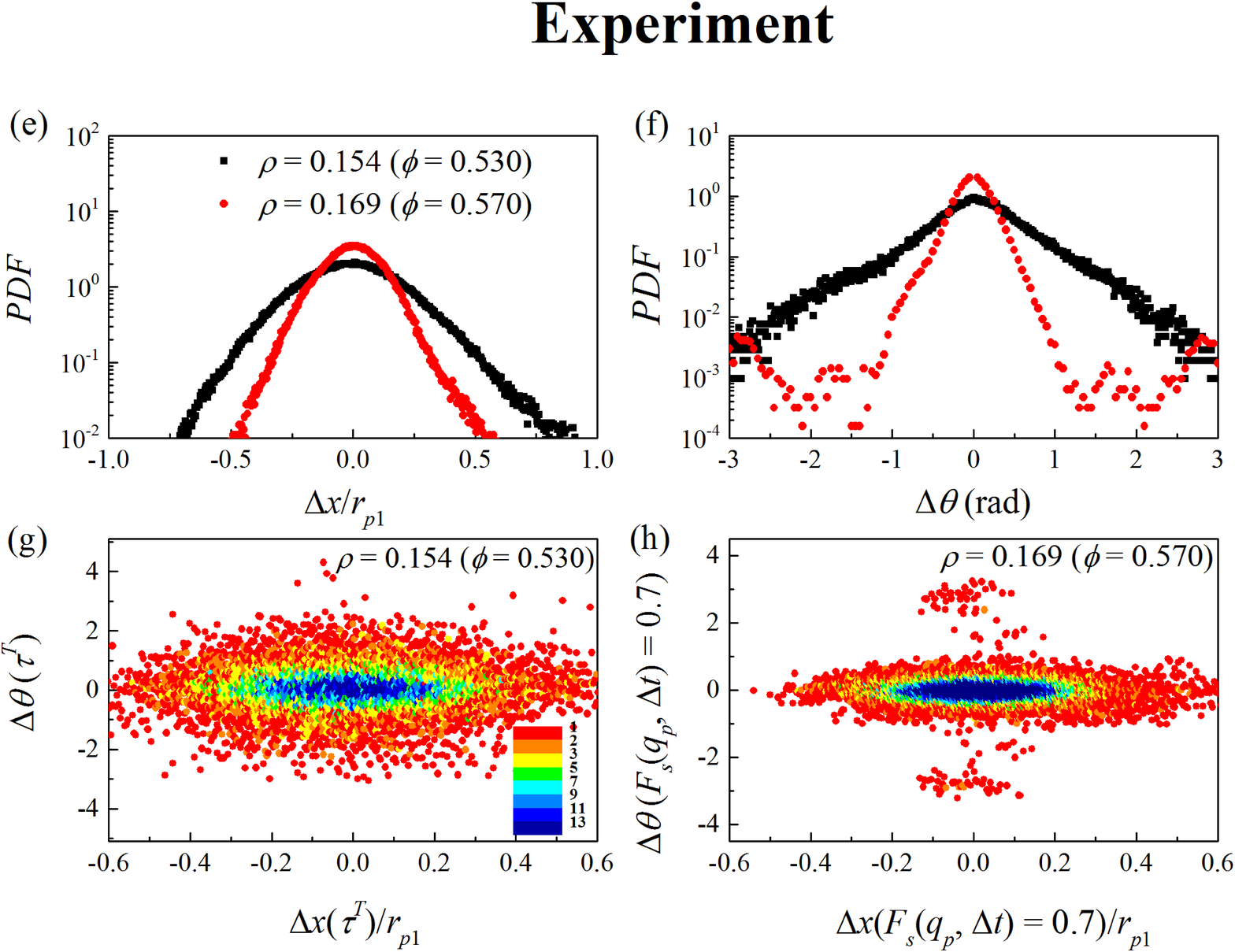}
 \caption{Probability distribution functions of translational ((a) and (e)) and rotational ((b) and
(f)) displacements, and correlation map of translational and
rotational displacements ((c), (d), (g) and (h)).
The top panel illustrates numerical data, the bottom panel experimental one,
in both cases at $\rho=0.154$ ($\phi=0.530$), in the liquid regime, and at $\rho=0.169$ ($\phi=0.570$), in the supercooled regime.
The displacements are evaluated after at
$\Delta t=\tau^{T}$ for $\rho=0.154$, and at the $\Delta t$ at which $F_{s}(q_{p},\Delta t)=0.7$ for $\rho=0.169$.
The translational displacements are scaled by the first peak position $r_{p1}$ of the corresponding pair correlation function.
}
  \label{fig:dtr}
\end{figure}

We relate the heterogeneous dynamics to the single particle motion investigating
the translational and the rotational displacement probability distribution functions (PDF). We discuss in this respect numerical results,
as well as experimental ones obtained by analyzing the experimental data of Ref.~\citenum{Kun}.
In Fig.~\ref{fig:dtr}, we show results for two values of density,
one in the normal liquid regime, $\rho=0.154$ ($\phi=0.530$),
and another in the supercooled regime, $\rho=0.169$ ($\phi=0.570$) (see Figs.~\ref{fig:dtr}(a,b) and (e,f)).
The low density displacements are evaluated in the time interval $\Delta t=\tau^{T}$,
while the high density data are calculated at the $\Delta t$ at which $F_{s}(q_{p},\Delta t)=0.7$, which is the minimum value of $F_s$ experimentally reached at high density.

In Figs.~\ref{fig:dtr}(a) (simulations) and \ref{fig:dtr}(e) (experiments) we observe that, as the system is compressed, the translational displacement PDF
increasingly deviates from a Gaussian distribution due to the emergence of exponential-like tails, in agreement with earlier reports~\cite{Weeks_Science}.
Conversely, Figs.~\ref{fig:dtr}(b) (simulations) and \ref{fig:dtr}(f) (experiments) show that the rotational displacement PDF deviate from a Gaussian also because of the emergence of secondary peaks. 
These peaks correspond to rotations in the range
$\Delta\theta \simeq 114^{\circ}:170^{\circ}$ ($2\sim3$ in radian unit).
This range is centered around the largest angle of the particles ($144^\circ$), and does not depend on the $\Delta t$ at which the displacement distribution is evaluated (see Fig.~\ref{fig:Dis} in Appendix~\ref{sec:Dhs}, where the same plot as Fig.~\ref{fig:dtr}(h) is illustrated for $\Delta t=\tau^{T}$), which indicates that the bumps are related to the particle shape. 

In Figs.~\ref{fig:dtr}(c), \ref{fig:dtr}(g) and \ref{fig:dtr}(d), \ref{fig:dtr}(h), we also illustrate the scatter plot of the single particle
rotational vs. translational displacement, respectively at small and at high densities.
At high density the scatter plot develops clouds corresponding to particles with large rotational displacements, which are those contributing to the bumps in the PDF.
These numerical and experimental results reveal that the rotational and
the translational displacements are uncorrelated.
This is also confirmed by the direct visualization of the particle trajectories, as shown in Fig.~\ref{fig:tra} in Appendix~\ref{sec:Dhs}.

The non-Gaussianity of the displacement distribution can be quantified by
the maximum peak heights of the translational ($\alpha_{2}^{p,T}$) and rotational
($\alpha_{2}^{p,R}$) non-Gaussian parameters.
We find $\alpha_{2}^{p,R}$ to be more than
ten times larger than $\alpha_{2}^{p,T}$ in the deep supercooled regime (see Fig.~\ref{fig:ngp} in Appendix~\ref{sec:Dhs}), in agreement with previously reported experimental findings~\cite{Kun}.
Since particles with larger displacements contribute more to
the non-Gaussian parameter, the large value of
$\alpha_{2}^{p,R}$ is clearly related to the observed bumps in the rotational displacement distribution function.

To explore the link between the non-Gaussian behavior
and the breakdown of the rotational inverse proportionality between diffusion coefficient and relaxation time shown in Fig.~\ref{fig:se},
we consider that for small angular displacements the $PDF(\Delta\theta)$
is well approximated by a Gaussian function~\cite{Granick_Gaussian},
as we verified by checking the existence of a linear
relation between $\log(PDF(\Delta\theta)/PDF(0))$ and $\Delta\theta^2$.
A Gaussian fit restricted to the range
of $\Delta\theta$ where this linear relationship holds
allows to extract characteristic diffusion coefficients for the slow particles,
$D^{R,slow}$ and $D_{CR}^{R,slow}$,
which are also plotted as a function of the relaxation time
in Figs.~\ref{fig:se}(a) and~\ref{fig:se}(b), respectively.
We observe that $D^{R,slow}$ is not inverse proportional to $\tau^T$,
due to the effect of the long-wavelength fluctuation.
Conversely, $D_{CR}^{R,slow} \propto 1/\tau_{CR}^T$.
Since the breakdown of the inverse proportionality is seen in $D_{CR}^{R}$ vs. $\tau_{CR}^T$, but not in $D_{CR}^{R,slow}$ vs. $\tau_{CR}^T$, we conclude that this breakdown is due to the particles with a large rotational displacement.

\subsection{Spatial correlation and length scales}
\begin{figure}[!tb]
 \centering
 \includegraphics[angle=0,width=0.45\textwidth]{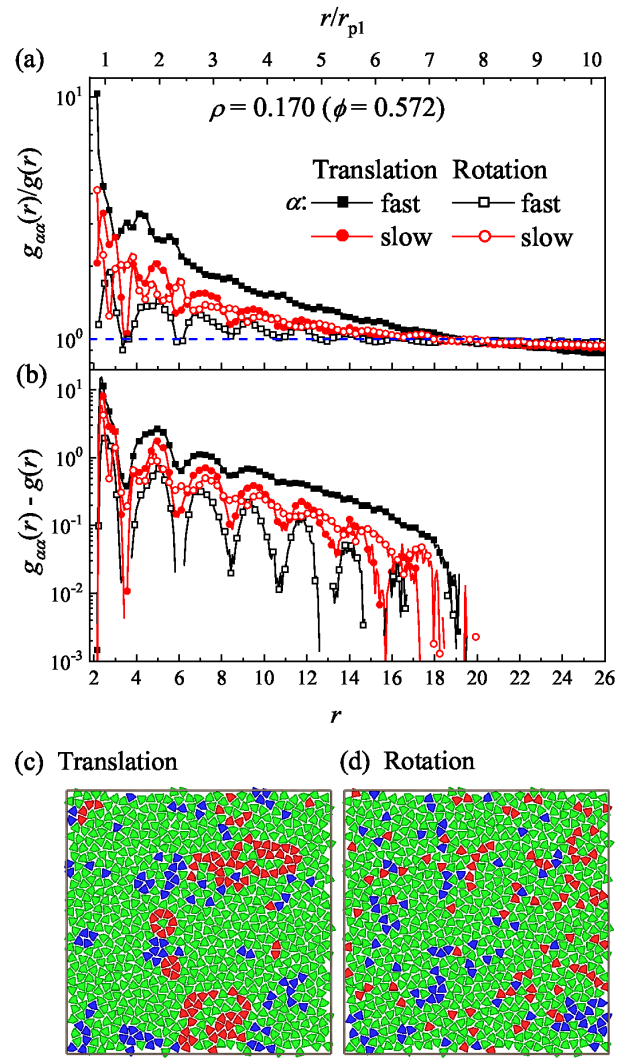}
 \caption{$r$ dependence of (a) $g_{\alpha\alpha}(r)/g(r)$ and of (b) $g_{\alpha\alpha}(r)-g(r)$, where
$g_{\alpha\alpha}(r)$ is the pair correlation function between fast (black) and between slow (red) translational (solid symbols) or rotational (open symbols) particles. \textcolor{black}{The length scales in (a) and (b) are expressed in bare unit on the bottom axis and are scaled by one typical particle size, which is evaluated by the first peak position $r_{p1}$ of pair correlation functions, on the top axis.}
(c) and (d) are snapshots of the system, where red (blue) particles represent the top $10\%$ fastest (slowest) ones in translation (c) and in rotation (d) during the time interval $\tau^{T}$.}
  \label{fig:gr}
\end{figure}

\textcolor{black}{
The absence of a clear correlation between the translational and the rotational motion (Fig.~\ref{fig:dtr}) suggests that the translational and the rotational degrees of freedom might relax through different processes.
These relaxation processes could be unveiled investigating the correlation lengths associated with the translational and rotational dynamical heterogeneities.
Previous works~\cite{DH_book} have shown that the dynamical length scale quantifying the spatial correlation of the fast (or slow) particles grows in the supercooled regime possibly diverging close to the MCT glass transition point~\cite{DH_book,Tanaka2015PNAS,Szamel_2d3d}, although 
a non-monotonic change of this length scale has also been reported~\cite{Kobnature,HimaNagamanasa2015}.
The relation between this dynamical length scale and static length scales associated with the spatial correlation of structural motifs is debated.
For instance, the static hexatic length scale grows as the dynamical length scale increases and diverges at the MCT point, in polydisperse discs~\cite{Tanaka2008PRL, Tanaka2011_natureMaterial}.
Differently, the amorphous length scale quantified by the point-to-set length scale only grows mildly in the supercooled region~\cite{Biroli2007Np,Kobnature,Tanaka2015PNAS}, possibly diverging only at zero temperature~\cite{Berthier_2019nc}.
}

Here we investigate the spatial correlation functions $g_{\alpha\alpha}(r)$ of
translational and rotational fast ($\alpha=$ fast) and slow ($\alpha=$ slow)
particles~\cite{Kob_1997}. 
Fast (slow) particles are defined as particles
with top $10\%$ largest (smallest) values of displacement within $\tau^{T}$ in
translation or in rotation. 
\textcolor{black}{We compare $g_{\alpha\alpha}(r)$ with the bulk pair
correlation function $g(r)$ by calculating $g_{\alpha\alpha}(r)/g(r)$ and $g_{\alpha\alpha}(r) -g(r)$.
If the fast (slow) particles are randomly distributed throughout the system, then $g_{\alpha\alpha}(r)/g(r)=1$ and $g_{\alpha\alpha}(r) -g(r)=0$.
Figures~\ref{fig:gr}(a) and \ref{fig:gr}(b) illustrate that $g_{\alpha\alpha}(r)$
is actually larger than $g(r)$ at short distances, and smaller than $g(r)$ at long ones.}
This indicates that the selected fast (slow) particles are clustered.
The figures also reveal that correlations in the translational motion are stronger
than correlations in the rotational one.
This result is consistent with the direct visualization of translational and
rotational fast (blue) and slow (red) particles shown in Figs.~\ref{fig:gr}(c) and~\ref{fig:gr}(d).
Indeed, clusters formed by translational fast particles are clearly
larger than those formed by rotational fast particles, indicating that the translational
relaxation is much more cooperative than the rotational one.
This possibly occurs as the convex shape of the kites allow
a particle to rotate without a strong cooperative motion of neighboring particles. \textcolor{black}{This mechanism could also lead to the decoupling between translational and rotational diffusion, as previously observed in Brownian squares~\cite{Kun_squares} and pentagons~\cite{Kun_Pentagons} and in a system of spherical tracers in a colloidal suspensions~\cite{Granick_tracer}.}
We therefore speculate that a stronger correlation between the rotational and the translational motion, e.g., without breakdown of the rotational inverse proportionality between diffusion coefficient and relaxation time, might occur in systems of non-convex polygons, such as star-shaped polygons~\cite{Mason_book}.

\begin{figure}[!tb]
 \centering
 \includegraphics[angle=0,width=0.45\textwidth]{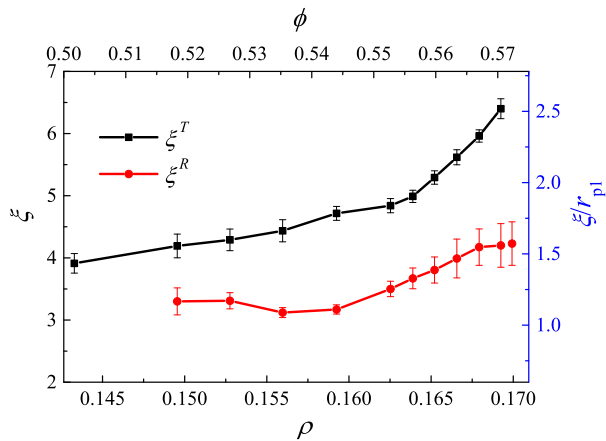}
 \caption{$\rho$ (or $\phi$) dependence of the dynamical correlation lengths for
both translation (black squares) and rotation (red circles). \textcolor{black}{The length scales are expressed in bare unit on the left black axis and are scaled by one typical particle size $r_{p1}$, on the right blue axis.}}
  \label{fig:length}
\end{figure}

The extent to which particle motion is correlated can be measured investigating
the dynamical correlation length associated to the translational and rotational motion, respectively.
To extract these two length scales we investigate the spatio-temporal correlation function~\cite{Pastore}:
\begin{equation}
g_{4}(r,t)=\langle\omega_{i}(t)\omega_{j}(t)\rangle -
\langle\omega_{i}(t)\rangle \langle\omega_{j}(t)\rangle.
\label{eq:g4}
\end{equation}
Here $r=|\mathbf{r}_{i}(0)-\mathbf{r}_{j}(0)|$ and $\omega_{i}(t)=1(0)$ if
$|\mathbf{r}_{i}(t)-\mathbf{r}_{i}(0)|\leq$ ($>$) $l_{*}^{T}$, for
translation and if $|\theta_{i}(t)-\theta_{i}(0)|\leq$ ($>$) $l_{*}^{R}$,
for rotation. We fix the threshold values $l_{*}^{T} = 0.8$ and $l_{*}^{R} = 0.25$,
the values at which the peak height of the corresponding translational and rotational
four-point susceptibility $\chi_{4}(t)=1/N\sum_{i,j}g_{4}(r,t)$ is maximal, at $\rho=0.169$ ($\phi=0.570$).
Figure~\ref{fig:g4rt} in Appendix~\ref{sec:stcf} shows that the corresponding
spatial-temporal correlation functions~\cite{Pastore} decay exponentially.
This allows to extract a translational and rotational dynamical length scale.
The maximum values of these length scales,
$\xi^{T}(\rho)$ and $\xi^{R}(\rho)$, that are respectively attained
at time $\tau^{T}(\rho)$ and $\tau^{R}(\rho)$, increases as the system is compressed, as illustrated in Fig.~\ref{fig:length}.
In line with the results of Fig.~\ref{fig:gr} we observe $\xi^{T} > \xi^{R}$.
The two length scales are clearly different and not proportional,
as $\xi^{T}$ grows steadily while $\xi^{R}$ grows slowly and possibly saturates at large densities.
A similar result is obtained investigating analogous CR length scale.
The different area fraction dependence of these length scales indicates
that the translational and the rotational degrees of freedom relax via different physical
processes.

\section{Conclusions}
We have performed a numerical investigation of the glassy dynamics of a 2D system of crowded monodisperse Penrose kites, and provided some supporting experimental measures. 
We have first investigated the melting transition of this system, which occurs through a first-order solid-liquid transition, and have then shown that upon compression the system reaches a supercooled (overcompressed) glassy state with no signature of nematic or crystalline order.
The investigation of the glassy dynamics revealed two important features, as described below.

First, we have clarified that long-wavelength fluctuations affect the scaling relation between diffusion coefficient and relaxation time \textcolor{black}{in our 2D system},
as clear from the comparison of the standard and of the CR measures.
In particular, our results show that the surprising but ubiquitous~\cite{SE_softdisk,Sastry_SE,SE_Ellipses, Sung_Tracer} breakdown of the inverse proportionality between diffusion coefficient and relaxation time in the normal liquid regime has to be attributed to these long-wavelength fluctuations.
Indeed, those inverse proportionalities in both translation and rotation are verified in the normal liquid regime by evaluating dynamics using CR quantities that filter out collective particle displacements.
In the supercooled regime the translational diffusion coefficient keeps its inverse proportionality to the relaxation time when evaluated by CR measures: this clarifies that the CR measures, by suppressing the effect of correlated particle displacements regardless of their physical origin, is less affected by the dynamical heterogeneities~\cite{unpublished}.
Care should therefore be taken when using CR measures in the investigation of the supercooled dynamics, as there is a risk of throwing the baby (i.e., dynamical heterogeneities) out with the bath water (i.e., long-wavelength fluctuations).
Conversely, the inverse proportionality between the rotational diffusion coefficient and the relaxation time exhibits a standard breakdown in the supercooled regime, which indicates that the rotational motion is mostly temporal rather than spatial heterogeneous.

Secondly, we have shown both experimentally and numerically that the translational and the rotational motion are uncorrelated and characterized by different dynamical correlation lengths. We have extracted the typical length scales, and found that the translational motion is more highly correlated in space than the rotational one.
The absence of correlation between translation and rotation might be attributed to the shape of the particles, which allows particles to rotate with little disturbance of neighboring particles.

Overall, our results provide insights into the rotational and translational glassy dynamics in a system with polymorphic local arrested structures. 
We have revealed the unexpected scaling between diffusion coefficient and relaxation time in both translation and rotation in 2D and the relationship between long-wavelength fluctuations and dynamical heterogeneities. 
These findings, in turn, will likely stimulate further research on the interesting and complex relationship between translational motion, rotational motion and particle shape.

\section*{Acknowledgement}
MPC and YWL acknowledge support from the Singapore Ministry of Education
through the Academic Research Fund (Tier 2) MOE2017-T2-1-066 (S) and from the National
Research Foundation Singapore, and are grateful to the National Supercomputing Centre (NSCC)
of Singapore for providing computational resources.
KZ acknowledges the support from the National Natural Science Foundation of China (21573159 and 21621004).
ZYS acknowledges the support from the National Natural Science Foundation of China (21833008, 21790344)
and the Key Research Program of Frontier Sciences, CAS (QYZDY-SSW-SLH027).
TGM acknowledges financial support from UCLA.

\appendix

\section{Mapping experiments and simulations\label{sec:simexp}}

\subsection{Pair correlation functions\label{sec:Mapsize}}
\begin{figure}[b!]
 \centering
 \includegraphics[angle=0,width=0.42\textwidth]{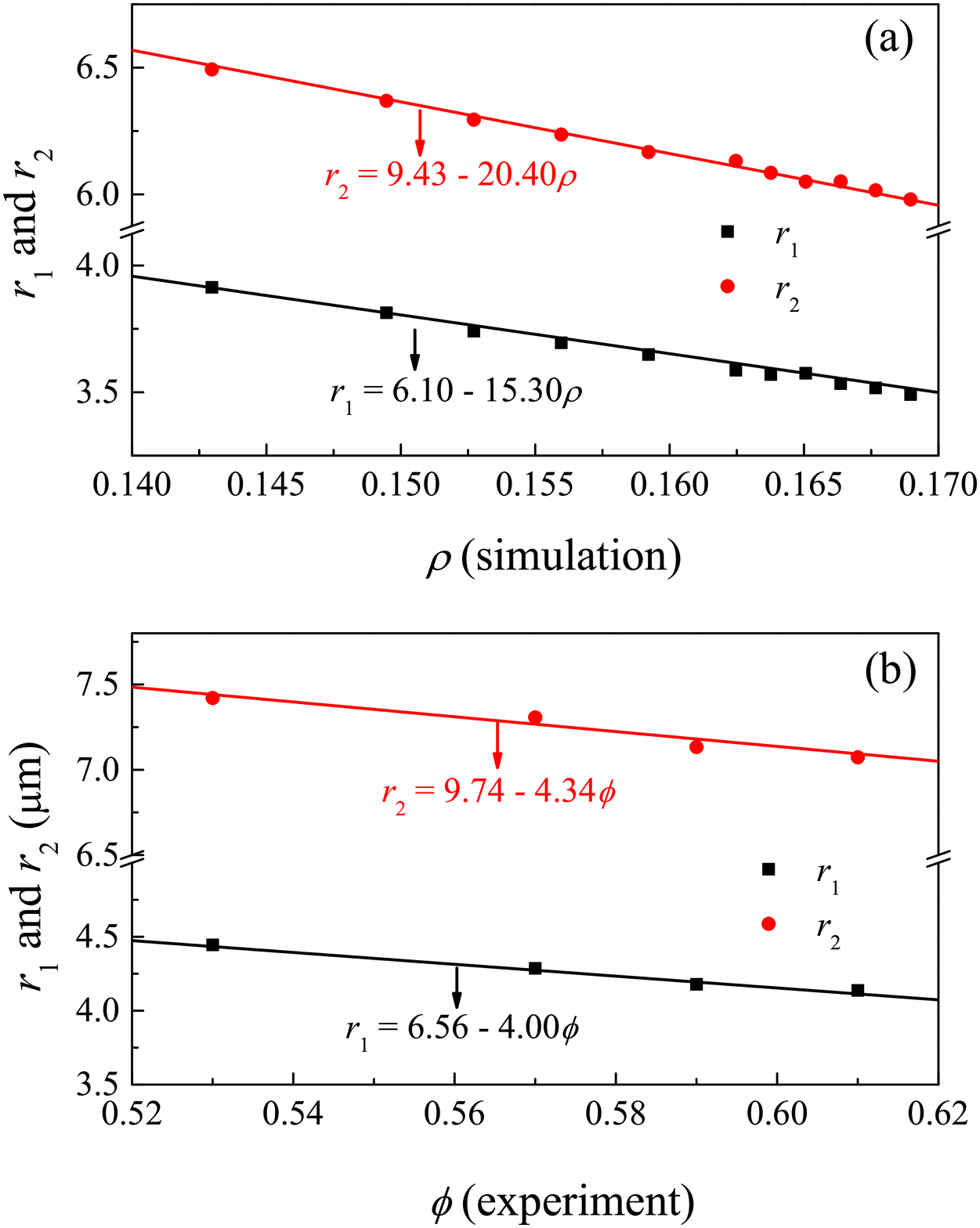}
 \caption{
 Dependence of position of the first $r_{1}$ and of the second $r_{2}$ minimum of the pair correlation functions $g(r)$
 on the number density $\rho$ for the simulation (a), and on the area fraction $\phi$ for experiments (b).
 Lines are linear regression fits.}
  \label{fig:r1r2}
\end{figure}

\begin{figure}[t!]
 \centering
 \includegraphics[angle=0,width=0.42\textwidth]{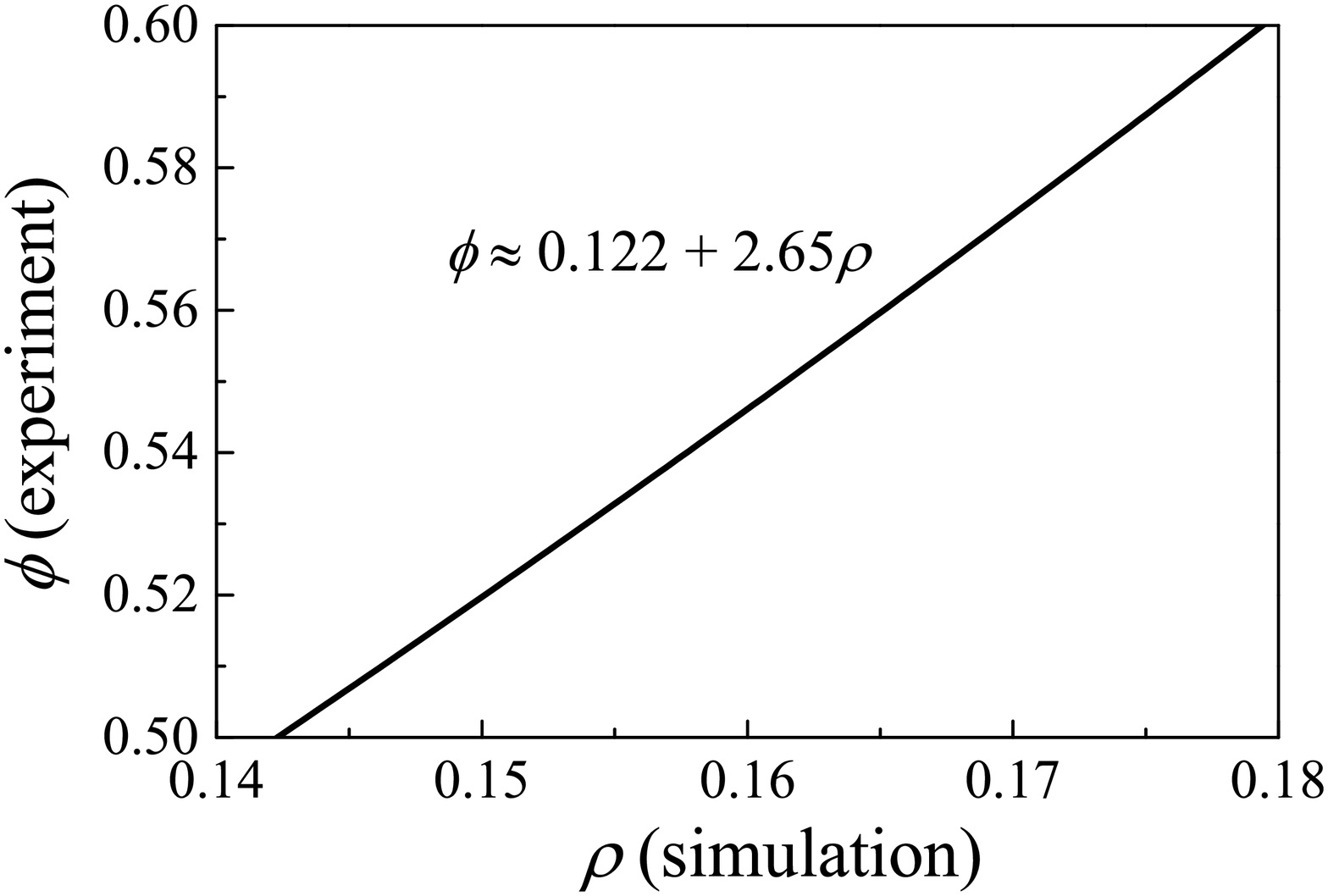}
 \caption{Mapping the numerical number density ($\rho$) to the experimental packing fraction ($\phi$). A linear function, $\phi \simeq 0.122+2.65\rho$, well captures the relationship between $\rho$ and $\phi$.}
  \label{fig:size}
\end{figure}

In the numerical model, a kite is represented by a collection of rigidly connected point particles. These particles interact with particles of other kites via a soft potential.
Because of this, the definition of kite size, and hence of volume fraction, is not straightforward.
To compare with the experiments, here we assume that an experimental and a numerical systems have the same area fraction when they have the same $r_2/r_1$ ratio, where $r_1$ and $r_2$ are the positions of the first two minima of the pair correlation function. 
We locate $r_1$ and $r_2$ positions by performing local fits of $g(r)$ with a Gaussian function.
In Fig.~\ref{fig:r1r2}, we illustrate their dependence on the number density $\rho$, for the simulations, and on $\phi$, for the experiments.
Both $r_{1}$ and $r_{2}$ decrease linearly as $\rho$ or $\phi$ increases, resulting in two linear functions $r_{1}=a_{1} + b_{1}\rho$ and $r_{2}=a_{2} + b_{2}\rho$, respectively. From these linear fits, we estimate the concentration dependence of $r_2/r_1$. \textcolor{black}{Figure~\ref{fig:size} shows the dependence of $\phi$, for the experiments, on $\rho$, for the simulations. In the investigate density range, the liner function $\phi\simeq0.122+2.65\rho$ captures the relationship between $\rho$ and $\phi$.
}

As an example of the validity of the above approach, we compare in Fig. 1(b) the experimental and the numerical pair correlation functions at $\rho=0.154$ ($\phi=0.530$) and at $\rho=0.169$ ($\phi=0.570$). The numerical system mimics the experimental one well in pair structure.

\subsection{The relative pointing angle between a kite and its closest neighbor\label{sec:Rangle}}

\begin{figure}[t!]
 \centering
 \includegraphics[angle=0,width=0.46\textwidth]{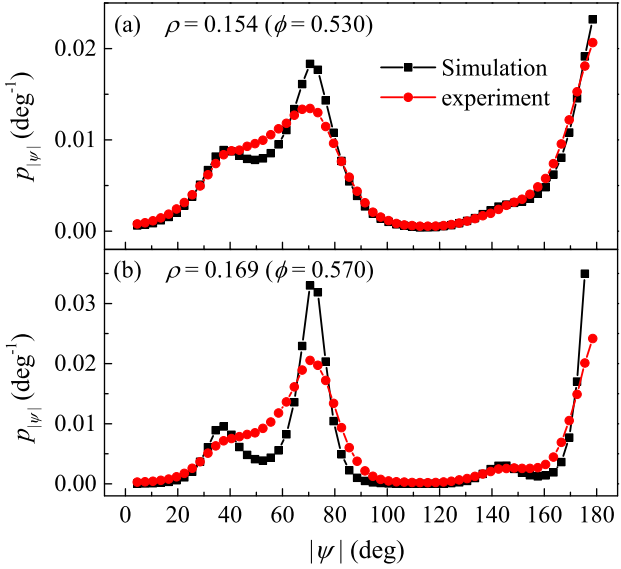}
 \caption{Probability distribution of relative pointing angle $\psi$ between a kite and its closest neighbor. We show the data from simulation (black squares) and experiment (red circles) at (a) $\rho=0.154$ ($\phi=0.530$) and at (b) $\rho=0.169$ ($\phi=0.570$).}
  \label{fig:angle}
\end{figure}

\begin{figure}[b!]
 \centering
 \includegraphics[angle=0,width=0.40\textwidth]{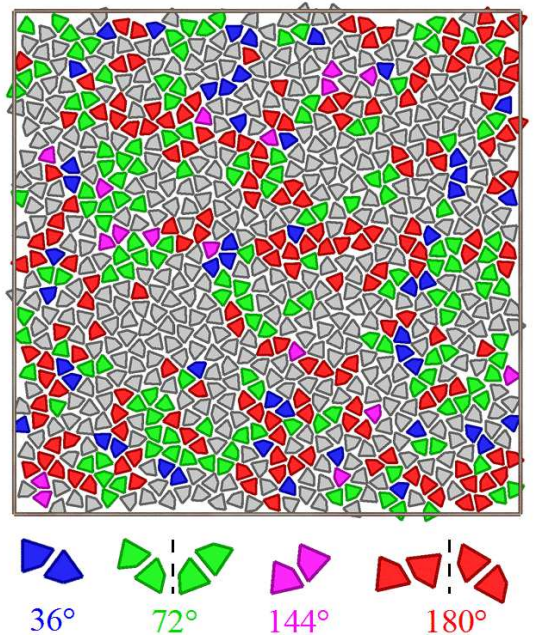}
 \caption{Visualization for the spatial distribution of $\psi$ for $\rho=0.169$ ($\phi=0.570$). The color code is listed at the bottom except that the gray color represents kites whose $\psi$ do not correspond to any peak (or bump) in $p_{|\psi|}$ illustrated in Fig.~\ref{fig:angle}}.
  \label{fig:Snapshot}
\end{figure}

As an additional check of the validity of our approach to map simulations and experiments, we compare in Fig.~\ref{fig:angle} the experimental and numerical distributions of the relative pointing angle $\psi$ between a kite and its closest neighbor. 
We observe two main peaks at $72^{\circ}$ and $180^{\circ}$ and two secondary peaks at $36^{\circ}$ and $144^{\circ}$. 
The secondary peaks become more prominent on increasing $\rho$ ($\phi)$. 
Numerical data well reproduces the experimental ones. They overestimate the height of the peaks, consistently with the observed higher heights of the first four peaks of the numerical pair correlation function.
This suggests that in the numerical model kites pack slightly better than in the experiments, possibly because of the absence of any polydispersity.

Figure~\ref{fig:Snapshot} depicts a snapshot of the system 
at $\rho=0.169$ ($\phi=0.570$), with color codes associating to $|\psi|$. 
We notice that the kites with $|\psi|$ corresponding to the dominant $72^{\circ}$ and $180^{\circ}$ form small clusters, as experimentally observed~\cite{Kun}.

\section{Translational and rotational dynamics}
\subsection{Dynamical arrest\label{sec:dyna}}
\begin{figure}[b!]
 \centering
 \includegraphics[angle=0,width=0.40\textwidth]{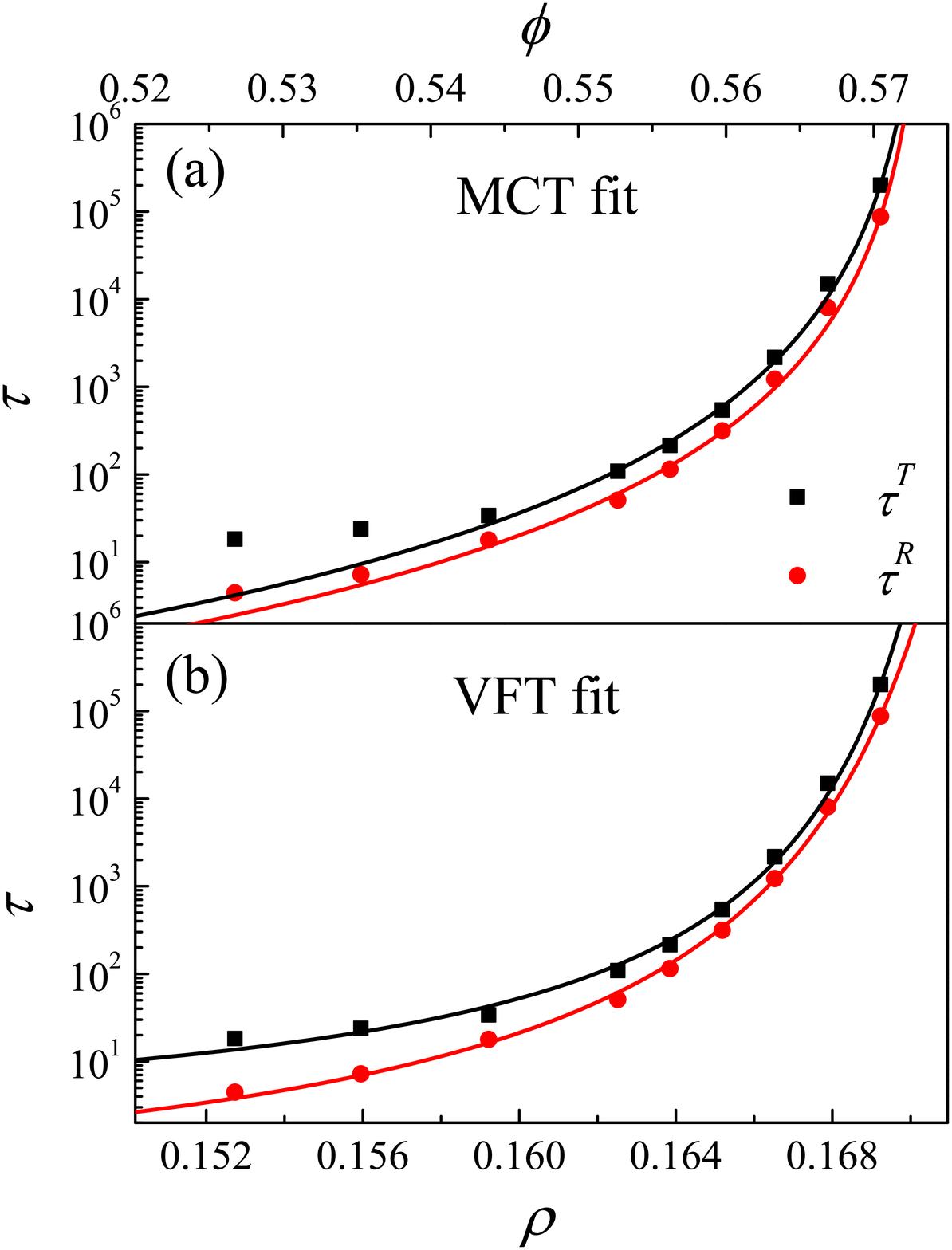}
 \caption{$\rho$ ($\phi$) dependence of translational $\tau^{T}$ (solid squares) and rotational $\tau^{R}$ (red circles) relaxation time. Lines in (a) are from MCT fitting whereas lines in (b) are from VFT fitting.}
  \label{fig:fit}
\end{figure}

We have checked whether the slowing down of the dynamics can be described by the power-law prediction of the mode-coupling theory (MCT), $\tau \sim(\rho_{c}-\rho)^{-\gamma}$, and by the Vogel-Fulcher-Tammann (VFT) law, $\tau \sim \exp(D_{f}\rho/(\rho_{0}-\rho))$. Here,  $\rho_{c}$ and $\rho_{0}$ mark the mode-coupling critical point and ideal glass transition point, respectively. $\gamma$ is a system constant and $D_{f}$ is the fragility parameter. Smaller value of $D_{f}$ corresponds to a steeper increase of $\tau$ as $\rho\rightarrow\rho_{0}$, resulting in a more fragile glass former. We present the fitting results in Fig.~\ref{fig:fit}. We find the MCT fitting parameters $\rho_{c}^{T}=0.171\pm0.0008$ and $\gamma^{T}=4.117\pm0.01$ for translation and $\rho_{c}^{R}=0.171\pm0.0013$ and $\gamma^{R}=4.001\pm0.023$ for rotation, and the VFT fitting parameters $\rho_{0}^{T}=0.174\pm0.0008$ and $D_{f}^{T}=0.238\pm0.004$ for translation and $\rho_{0}^{R}=0.174\pm0.0002$ and $D_{f}^{R}=0.361\pm0.006$ for rotation. Note that the obtained values of $\gamma$ and $D_{f}$ are all comparable to that of other glass models, e.g., Kob-Andersen model ~\cite{KA_95} or polydisperse Weeks-Chandler-Andersen (WCA) disks ~\cite{Kawasaki_polydiperse}, indicating that the kite system has similar nature of glass transition as other glass models both in translation and in rotation. The translational dynamics is more fragile than the rotational one since $D_{f}^{T}<D_{f}^{R}$.

\subsection{Different dynamical regimes\label{sec:regime}}

\begin{figure}[htb]
 \centering
 \includegraphics[angle=0,width=0.41\textwidth]{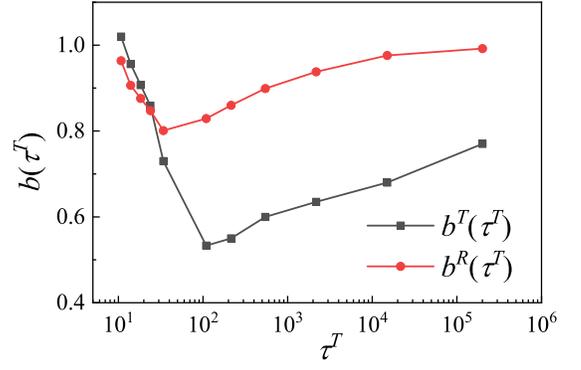}
 \caption{$\tau^{T}$ dependence of derivatives of the logarithm of MSD  $b^{T}(\tau^{T})$ (black squares) and of the logarithm of MSAD $b^{R}(\tau^{T})$ (red circles) at time $\tau^{T}$.
 }
  \label{fig:delta}
\end{figure}

As time advances, a supercooled liquid traverses different sub-diffusive regimes, identified by the logarithmic slope of the MSD
$b^{T}(t)=\mathrm{d}\left(\mathrm{ln}\langle \Delta r ^{2}(t)\rangle\right)/\mathrm{d(\ln(t))}$, and by that of the rotational MSAD, $b^{R}(t)$ (see Figs. 2(b) and 2(d)). 
At short times $b \simeq 1$ (overdamped dynamics) or $b \simeq 2$ (underdamped dynamics), in the intermediate regime where particles are confined by their neighbors $b < 1$, while $b = 1$ in the asymptotic diffusive regime.
To determine in which regime the system relaxes, we illustrate in Fig.~\ref{fig:delta} the $\tau^T$ dependence of
$b^{T}(\tau)$ and of $b^{R}(\tau)$. Both quantities have a non-monotonic dependence on the translational relaxation time.

\subsection{Dynamic heterogeneities\label{sec:Dhs}}
\begin{figure}[t!]
 \centering
 \includegraphics[angle=0,width=0.47\textwidth]{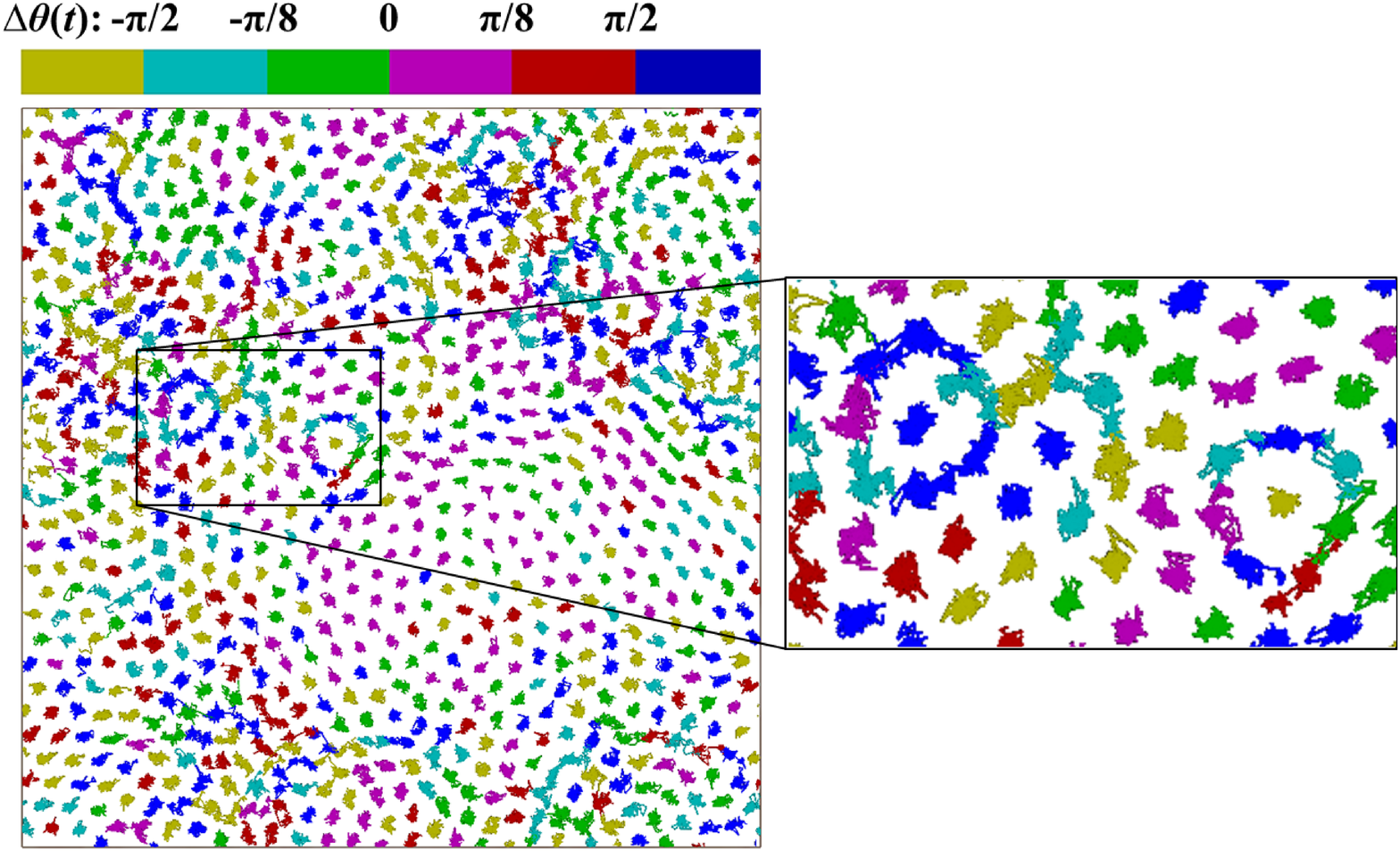}
 \caption{Translational trajectories of kites, with the color codes corresponding to the rotational displacements during an interval of $1.5\tau^{T}$ at $\rho=0.170$ ($\phi=0.572$). The enlargement shows the loop-like cooperative motion in translation.}
  \label{fig:tra}
\end{figure}

\begin{figure}[htb]
 \centering
 \includegraphics[angle=0,width=0.44\textwidth]{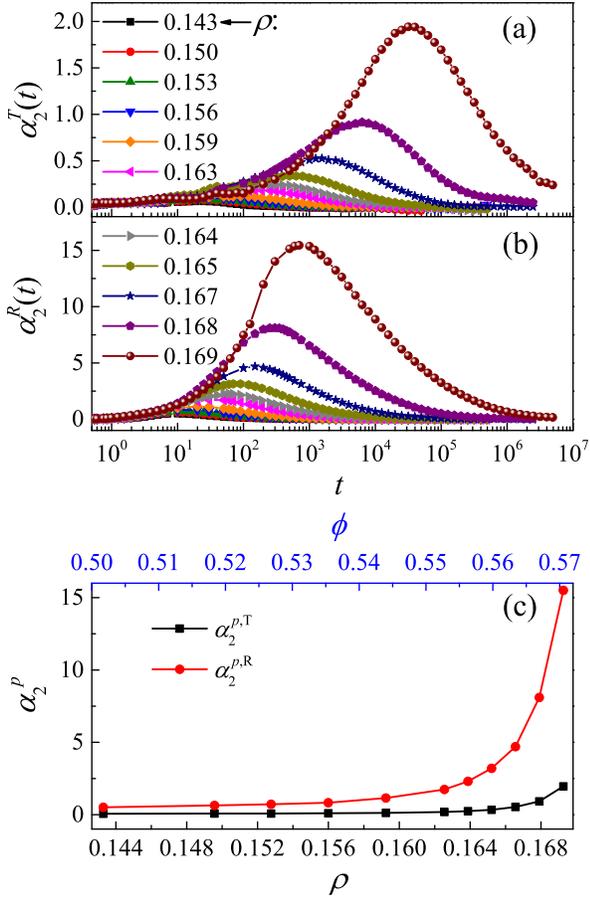}
 \caption{Time evolution of (a) translational and (b) rotational non-Gaussian parameter for different $\rho$, as shown in the legend. (c) Translational (black squares) and rotational (red circles) peak heights of the non-Gaussian parameter as a function of $\rho$ ($\phi$).}
  \label{fig:ngp}
\end{figure}

\begin{figure}[htb]
 \centering
 \includegraphics[angle=0,width=0.40\textwidth]{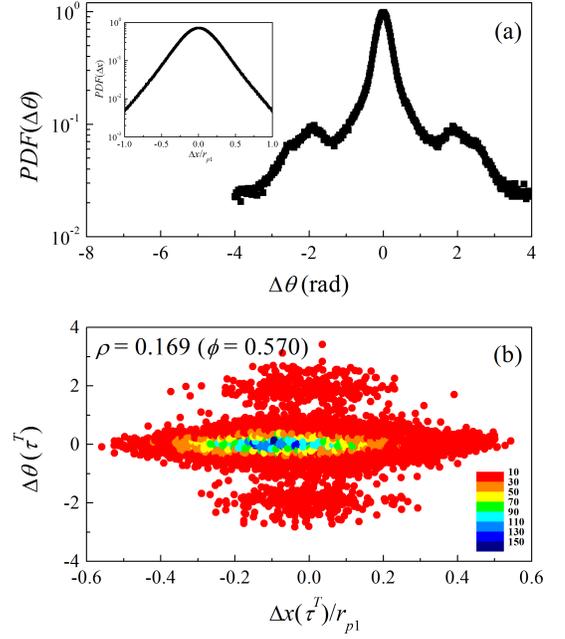}
 \caption{(a) Probability distribution functions of rotational and translational (inset) displacements in the time interval $\tau^{T}$ for $\rho=0.169$ ($\phi=0.570$). (b) The corresponding correlation map of translational and rotational displacements. The translational displacements are scaled by the first peak position $r_{p1}$ of the pair correlation function.}
  \label{fig:Dis}
\end{figure}

We visually illustrate the translational and rotational dynamic heterogeneities of the system in Fig.~\ref{fig:tra}, where the trajectories of different kites are colour coded according to the  magnitude of the rotational displacements.
In the figure, we see the emergence of a string-like trajectories,
and in a few cases the rotation of a group of kites around a central one (see the zoomed region in Fig.~\ref{fig:tra}).
There is not a clear correlation between the length of the trajectories and their colour, a further evidence of the decoupling of translational and rotational motion. This is consistent with the results of Fig.~6.

To quantify the degree of heterogeneity we resort to the non-Gaussian parameter (NGP), which measures how much the displacement distribution differ from the Gaussian behavior expected in a normal liquid.
The translational and rotational NGPs are defined as $\alpha_{2}^{T}(t)=\langle \Delta x^{4}(t)\rangle/3\langle \Delta x^{2}(t)\rangle-1$ and $\alpha_{2}^{R}(t)=\langle \Delta \theta^{4}(t)\rangle/3\langle \Delta \theta^{2}(t)\rangle-1$, respectively. Here, $\Delta x(t)$ is the displacement in $x$ coordinate within time $t$.
We find both $\alpha_{2}^{T}(t)$ and $\alpha_{2}^{R}(t)$ to develop pronounced peaks, with their peak times increasing as the dynamics slows down (see Fig.~\ref{fig:ngp} (inset)).
In addition, we note that $\alpha_{2}^{T}(t)$ reaches its peak value after $\alpha_{2}^{R}(t)$, which is in consistent with the fact that the translational relaxation time $\tau^{T}$ is larger than the rotational relaxation time $\tau^{R}$.
More interestingly, the peak height of $\alpha_{2}^{R}(t)$ is much larger than that of $\alpha_{2}^{T}(t)$ (see Fig.~\ref{fig:ngp}), especially at high area fraction.
This is so as $\alpha_{2}^{T}(t) > 0$ due to the emergence of exponential tails in the associated probability distribution, as in the inset of Fig.~\ref{fig:Dis}(a). Conversely, the rotational displacement distribution is highly non-Gaussian due to the presence of bumps at large rotational displacements, as in the main panel.
These bumps identify preferred values for the rotational displacement of the particles.
This is clear by comparing Fig.~\ref{fig:Dis}(a) and Fig.~6: in the two figures, which report the rotational displacement distribution at the same area fraction but at different times, the bumps occur at the same locations. Hence, the bumps do not diffuse. In the correlation map between the translational and the rotational displacements, Fig.~\ref{fig:Dis}(b), the bumps lead to the emergence of clusters of points at the preferred values of the rotational angle.

\section{Spatial-temporal correlation functions\label{sec:stcf}}
\begin{figure}[htb]
 \centering
 \includegraphics[angle=0,width=0.4\textwidth]{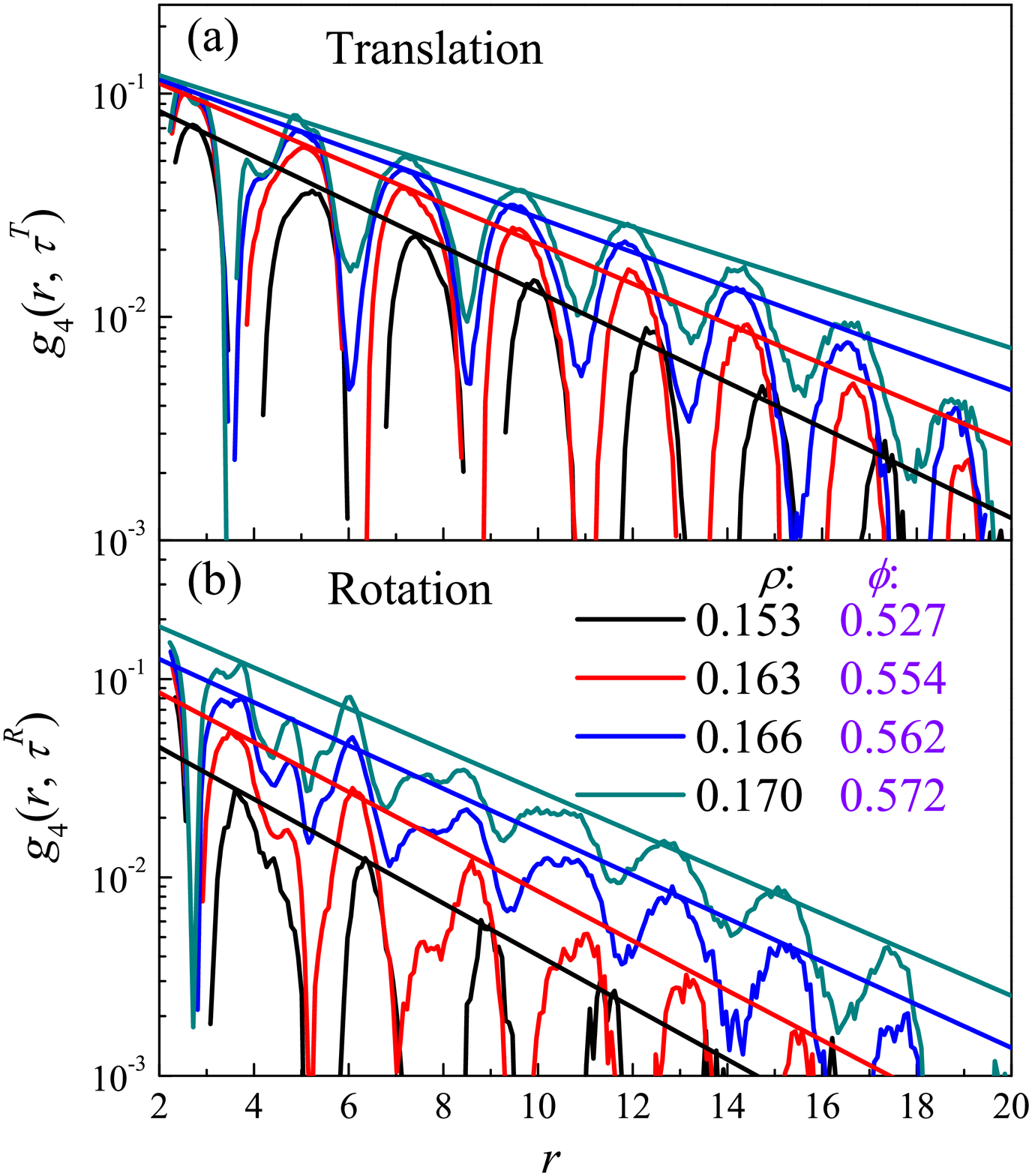}
 \caption{$\rho$ ($\phi$) dependence of the spatial-temporal correlation functions evaluated at time $\tau^{T}$ for translation (a) and at time $\tau^{R}$ for rotation (b). The solid lines are from exponential fittings $g_{4}(r,\tau)\sim \exp(-r/\xi)$.}
  \label{fig:g4rt}
\end{figure}

We have determined the dynamical correlation lengths associated to the relaxation of the system investigating the spatial-temporal correlation function $g_{4}(r,t)$. Specifically, since the temporal correlation is maximum at the relaxation time scale, we fix the time $t$ to be $\tau^{T}$ for translation and to be $\tau^{R}$ for rotation.
Figure~\ref{fig:g4rt} shows that these correlation functions decay exponentially with the distance, $g_{4}(r,\tau)\sim \exp(-r/\xi)$.
The dynamical length scales illustrated in Fig.~8 results from the illustrated exponential fits of these correlation functions.
\clearpage
\newpage


%
\end{document}